\def\w{\omega}

\def\xz{x^o}
\def\0{\emptyset}

\def\L{{\cal L}_{x^o}}
\def\FL{{\cal FL}_{x^o}}
\def\A{{\cal A}}
\def\G{{\cal G}}
\def\AG{\A/\G} \def\AGbar{\overline{\AG}}
\def\HA{{\cal HA}}
\def\HAbar{\overline{\cal HA}}
\def\HG{{\cal HG}{}}
\def\S{\Sigma}

\def\ut#1{\rlap{\lower1ex\hbox{$\sim$}}#1{}}
\def\At{\tilde{A}}  \def\Ab{\bar{A}}
\def\a{\alpha}  \def\at{\tilde\alpha}
\def\b{\beta}   \def\bt{\tilde\beta}
\def\g{\gamma}  \def\gt{\tilde\gamma}

\def\Tr{{\rm Tr}}
\def\C*{$C^\star$--algebra}
\def\Sstar{{\cal S}^\star}
\def\c{{\cal C}}
\def\cbar{\overline{\cal C}}
\def\T{{\cal T}}
\documentstyle{multi}
\begin{document}
\chapterb[Holonomy $C^*$ Algebras]{Representation Theory of Analytic
Holonomy $C^*$ Algebras}
{Abhay Ashtekar}
{Center for Gravitational Physics and Geometry, Penn
State University\protect\footnote
{The authors are grateful to John Baez for several illuminating
comments and to Roger Penrose for suggesting that the piecewise
analytic loops may be better suited for their purpose than the
piecewise smooth ones.  JL also thanks Clifford Taubes for stimulating
discussions. This work was supported in part by the National Science
Foundation grants PHY93--96246 and PHY91--07007; by the Polish
Committee for Scientific Research (KBN) through grant no.2 0430 9101;
and, by the Eberly research fund of Pennsylvania State University.}
%University Park, PA 16802--6300
%\centerline{\it ${}^1$Department of Physics, Syracuse University, Syracuse,
%N.Y. 13244--1130}
}
{Jerzy Lewandowski}
{Department of Physics, University of Florida
at Gainesville\protect\footnote{On leave from: Instytut Fizyki
Teoretycznej, Warsaw University}
}
{Ashtekar and Lewandowski}
\pagenumbering{arabic}
\medskip
\centerline{To appear in {\sl Knots and Quantum Gravity,}
        ed. J.Baez, Oxford U.Press}
\medskip
\begin{abstract}
Integral calculus on the space $\A/\G$ of gauge equivalent connections
is developed. Loops, knots, links and graphs feature prominently in
this description. The framework is well--suited for quantization of
diffeomorphism invariant theories of connections.

The general setting is provided by the abelian \C* of functions on
$\A/\G$ generated by Wilson loops (i.e., by the traces of holonomies
of connections around closed loops). The representation theory of this
algebra leads to an interesting and powerful ``duality'' between
gauge--equivalence classes of connections and certain equivalence
classes of closed loops. In particular, regular measures on (a
suitable completion of) $\A/\G$ are in 1--1 correspondence with
certain functions of loops and diffeomorphism invariant measures
correspond to (generalized) knot and link invariants. By carrying out
a non--linear extension of the theory of cylindrical measures on
topological vector spaces, a faithful, diffeomorphism invariant
measure is introduced. This measure can be used to define the Hilbert
space of quantum states in theories of connections. The Wilson--loop
functionals then serve as the configuration operators in the quantum
theory.
\end{abstract}

\section{Introduction}
The space $\A/\G$ of connections modulo gauge transformations plays an
important role in gauge theories, including certain topological
theories and general relativity [1]. In a typical set up, $\A/\G$ is
the classical configuration space. In the quantum theory, then, the
Hilbert space of states would consist of all square--integrable
functions on $\A/\G$ with respect to some measure. Thus, the theory of
integration over $\A/\G$ would lie at the heart of the quantization
procedure. Unfortunately, since $\A/\G$ is an infinite dimensional
{\it non--linear} space, the well--known techniques to define
integrals do not go through and, at the outset, the problem appears to
be rather difficult. Even if this problem could be overcome, one would
still face the issue of constructing self--adjoint operators
corresponding to physically interesting observables.  In a Hamiltonian
approach, the Wilson loop functions provide a natural set of
(manifestly gauge invariant) configuration observables. The problem of
constructing the analogous, manifestly gauge invariant ``momentum
observables'' is difficult already in the classical theory: these
observables would correspond to vector fields on $\A/\G$ and
differential calculus on this space is not well--developed.

Recently, Ashtekar and Isham [2] (hereafter referred to as A--I)
developed an algebraic approach to tackle these problems. The A--I
approach is in the setting of canonical quantization and is based on
the ideas introduced by Gambini and Trias in the context of Yang-Mills
theories [3] and by Rovelli and Smolin in the context of quantum
general relativity [4]. Fix a $n$--manifold $\S$ on which the
connections are to be defined and a compact Lie group $G$ which will
be the gauge group of the theory under consideration%
\footnote{ In typical applications, $\S$ will be a Cauchy surface in an
$n$+1 dimensional space--time in the Lorentzian regime or an
$n$--dimensional space--time in the Euclidean regime. In the main body
of this paper, $n$ will be taken to be $3$ and $G$ will be taken to be
$SU(2)$ both for concreteness and simplicity. These choices correspond
to the simplest, non--trivial applications of the framework to
physics.}.  The first step is the construction of a \C* of
configuration observables---a sufficiently large set of
gauge--invariant functions of connections on $\Sigma$. A natural
strategy is to use the Wilson--loop functions---the traces of
holonomies of connections around closed loops on $\Sigma$---to
generate this \C*. Since these are configuration observables, they
commute even in the quantum theory. The \C* is therefore Abelian. The
next step is to construct representations of this algebra. For this,
the Gel'fand spectral theory provides a natural setting since any
given Abelian \C* with identity is naturally isomorphic with the \C*
of continuous functions on a compact, Hausdorff space, the Gel'fand
spectrum $sp(C^\star)$ of that algebra.  (As the notation suggests,
$sp(C^\star)$ can be constructed directly from the given \C*: its
elements are homomorphism from the given \C* to the $\star$--algebra
of complex numbers.)  Consequently, every (continuous) cyclic
representation of the A--I \C* is of the following type: The carrier
Hilbert space is $L^2(sp(C^\star), d\mu)$ for some regular measure
$d\mu$ and the Wilson--loop operators act (on square--integrable
functions on $sp(C^\star)$) simply by multiplication. A--I pointed out
that, since the elements of the \C* are labelled by loops, there is a
1--1 correspondence between regular measures on $sp(C^\star)$ and
certain functions on the space of loops.  Diffeomorphism invariant
measures correspond to knot and link invariants.  Thus, there is a
natural interplay between connections and loops, a point which will
play an important role in the present paper.

Note that, while the classical configuration space is $\A/\G$, the
domain space of quantum states is $sp(C^\star)$. A--I showed that
$\A/\G$ is naturally embedded in $sp(C^\star)$ and Rendall [5] showed
that the embedding is in fact dense. Elements of $sp(C^\star)$ can be
therefore thought of as {\it generalized} gauge equivalent
connections. To emphasize this point and to simplify the notation, let
us denote the spectrum of the A--I \C* by $\AGbar$. The fact that the
domain space of quantum states is a completion of---and hence larger
than---the classical configuration space may seem surprising at first
since in ordinary (i.e. non--relativistic) quantum mechanics both
spaces are the same. The enlargement is, however, characteristic of
quantum field theory, i.e. of systems with an infinite number of
degrees of freedom. A--I further explored the structure of $\AGbar$
but did not arrive at a complete characterization of its elements.
They also indicated how one could make certain heuristic
considerations of Rovelli and Smolin precise and associate ``momentum
operators'' with (closed) strips (i.e.\ $n$--1 dimensional ribbons) on
$\Sigma$. However, without further information on the measure $d\mu$
on $\AGbar$, one can not decide if these operators can be made
self--adjoint, or indeed if they are even densely defined. A--I did
introduce certain measures with which the strip operators ``interact
properly.'' However, as they pointed out, these measures have support
only on finite dimensional subspaces of $\AGbar$ and therefore ``too
simple'' to be interesting in cases where the theory has an infinite
number of degrees of freedom.

In broad terms, the purpose of this paper is to complete the A--I program
in several directions. More specifically, we will present the following
results:

\begin{enumerate}
\item We will obtain a complete characterization of elements of the
Gel'fand spectrum. This will provide a degree of control on the domain
space of quantum states which is highly desirable: Since $\A/\G$ is a
non--trivial, infinite dimensional space which is not even locally
compact, while $\AGbar$ is a compact Hausdorff space, the completion
procedure is quite non--trivial. Indeed, it is the lack of control on
the precise content of the Gel'fand spectrum of physically interesting
$C^\star$ algebras that has prevented the Gel'fand theory from playing
a prominent role in quantum field theory so far.
\item We will present a faithful, diffeomorphism invariant measure on
$\AGbar$. The theory underlying this construction suggests how one
might introduce other diffeomorphism invariant measures. Recently,
Baez [6] has exploited this general strategy to introduce new
measures. These developments are interesting from a strictly
mathematical standpoint because the issue of existence of such
measures was a subject of controversy until recently%
\footnote{We should add, however, that in most of these debates,
it was $\A/\G$---rather than its completion $\AGbar$---that was the
center of attention.}.
They are also of interest from a physical viewpoint since one can,
e.g., use these measures to regulate physically interesting operators
(such as the constraints of general relativity) and investigate their
properties in a systematic fashion.
\item Together, these results enable us to define a Rovelli--Smolin
``loop--transform'' \ \ rigorously. \ \ This \ is \  a \ map \ from
 \ \ the Hilbert \ \ space

\noindent
$L^2(\AGbar, d\mu)$ to the space of suitably regular functions of loops
on $\S$. Thus, quantum states can be represented by suitable functions of
loops. This ``loop representation'' is particularly well--suited to
diffeomorphism invariant theories of connections, including general
relativity.
\end{enumerate}

We have also developed differential calculus on $\AGbar$ and shown that
the strip (i.e. momentum) operators are in fact densely--defined,
symmetric operators on $L^2(\AGbar, d\mu)$; i.e. that they interact
with our measure correctly. However, since the treatment of the strip
operators requires a number of new ideas from physics as well as certain
techniques from graph theory, these results will be presented in a
separate work.

The methods we use can be summarized as follows.

First, we make the (non--linear) duality between connections and loops
explicit. On the connection side, the appropriate space to consider is
the Gel'fand spectrum $\AGbar$ of the A--I \C*. On the loop side, the
appropriate object is the space of hoops---i.e., holonomically
equivalent loops. More precisely, let us consider based, piecewise
{\it analytic} loops and regard two as being equivalent if they give
rise to the same holonomy (evaluated at the base point) for {\it any}
($G$--)connection. Call each ``holonomic equivalence class'' a \-
($G$--)hoop. It is straightforward to verify that the set of hoops has
the structure of a group. (It also carries a natural topology [7],
which, however, will not play a role in this paper.) We will denote it
by $\HG$ and call it the {\it hoop group}. It turns out that $\HG$ and
the spectrum $\AGbar$ can be regarded as being ``dual'' to each other
in the following sense:

\begin{enumerate}
\item Every element of $\HG$ defines a continuous, complex--valued
function on $\AGbar$ and these functions generate the algebra of all
continuous functions on $\AGbar$;
\item Every element of $\AGbar$ defines a homomorphism from $\HG$ to
the gauge group $G$ (which is unique modulo an automorphism of $G$)
and every homomorphism defines an element of $\AGbar$.
\end{enumerate}

In the case of topological vector spaces, the duality mapping respects
the topology {\it and} the linear structure. In the present case,
however, $\HG$ has the structure only of a group and $\AGbar$ of a
topological space.  The ``duality'' mappings can therefore respect
only these structures.  Property 2 above provides us with a complete
(algebraic) characterization of the elements of Gel'fand spectrum
$\AGbar$ while property 1 specifies the topology of the spectrum.

The second set of techniques involves a family of projections from the
topological space $\AGbar$ to certain finite dimensional manifolds.
For each sub--group of the hoop group $\HG$, generated by $n$
independent hoops, we define a projection from $\AGbar$ to the
quotient, $G^n/{\rm Ad}$, of the n--th power of the gauge group by the
adjoint action. (An element of $G^n/{\rm Ad}$ is thus an equivalence
class of n--tuples, $(g_1, ...,g_n)$, with $g_i \in G$, where, $(g_1,
...,g_n) \sim (g_o^{-1} g_1 g_o, ...,g_o^{-1} g_n g_o)$ for any
$g_o\in G$.) The lifts of functions on $G^n/{\rm Ad}$ are then the
non--linear analogs of the cylindrical functions on topological vector
spaces. Finally, since each $G^n/{\rm Ad}$ is a compact topological
space, we can equip it with measures to integrate functions in the
standard fashion.  This in turn enables us to define cylinder measures
on $\AGbar$ and integrate cylindrical functions thereon. Using the
Haar measure on $G$, we then construct a natural, diffeomorphism
invariant, faithful measure on $\AGbar$.

Finally, for completeness, we will summarize the strategy we have
adop- ted to introduce and analyse the ``momentum operators,''
although, as mentioned above, these results will be presented
elsewhere. The first observation is that we can define ``vector
fields'' on $\AGbar$ as derivations on the ring of cylinder functions.
Then, to define the specific vector fields corresponding to the
momentum (or, strip) operators we introduce certain notions from graph
theory. These specific vector fields are ``divergence--free'' with
respect to our cylindrical measure on $\AGbar$, i.e., they leave the
cylindrical measure invariant. Consequently, the momentum operators
are densely defined and symmetric on the Hilbert space of
square--integrable functions on $\AGbar$.

The paper is organized as follows. Section 2 is devoted to
preliminaries.  In Section 3, we obtain the characterization of
elements of $\AGbar$ and also present some examples of those elements
which do not belong to $\A/\G$, i.e. which represent genuine,
generalized (gauge equivalence classes of) connections. Using the
machinery introduced in this characterization, in Section 4 we
introduce the notion of cylindrical measures on $\AGbar$ and then
define a natural, faithful, diffeomorphism invariant, cylindrical
measure. Section 5 contains concluding remarks.

In all these considerations, the piecewise analyticity of loops on
$\S$ plays an important role. That is, the specific constructions and
proofs presented here would not go through if the loops were allowed
to be just piecewise smooth. However, it is far from being obvious
that the final results would not go through in the smooth category. To
illustrate this point, in Appendix A we restrict ourselves to $U(1)$
connections and, by exploiting the Abelian character of this group,
show how one can obtain the main results of this paper using piecewise
$C^1$ loops. Whether similar constructions are possible in the
non--Abelian case is, however, an open question. Finally, in Appendix
B we consider another extension. In the main body of the paper, $\S$
is a 3--manifold and the gauge group $G$ is taken to be $SU(2)$. In
Appendix B, we indicate the modifications required to incorporate
$SU(N)$ and $U(N)$ connections on an n--manifold (keeping however,
piecewise analytic loops).

\section{Preliminaries}
In this section, we will introduce the basic notions, fix the notation
and recall those results from the A--I framework which we will need in
the subsequent discussion.

Fix a 3--dimensional, real, analytic, connected, orientable manifold
$\S$.  Denote by $P(\S, SU(2))$ a principal fibre bundle $P$ over the
base space $\S$ and the structure group $SU(2)$. Since any $SU(2)$
bundle over a 3--manifold is trivial we take  $P$ to be the product
bundle. Therefore, $SU(2)$ connections on $P$ can
be identified with $su(2)$--valued 1--form fields on $\S$. Denote by
$\A$ the space of smooth (say $C^1$) $SU(2)$ connections, equipped
with one of the standard (Sobolev) topologies (see, e.g., [8]). Every
$SU(2)$--valued function $g$ on $\S$ defines a gauge transformation in
$\A$,

\begin{equation}
A\cdot g:= g^{-1}Ag + g^{-1}dg.
\end{equation}

Let us denote the quotient of $\A$ under the action of this group $\G$
(of local gauge transformations) by $\A/\G$. Note that the affine
space structure of $\A$ is lost in this projection; $\A/\G$ is a
genuinely non--linear space with a rather complicated topological
structure.

The next notion we need is that of closed loops in $\S$. Let us begin
by considering continuous, piecewise analytic ($C^\w$) parametrized
paths, i.e., maps
\begin{equation}
p: [0, s_1]\cup ...\cup[s_{n-1}, 1]\rightarrow \S
\end{equation}
which are  continuous on the whole domain and $C^\w$  on the closed
intervals $[s_k, s_{k+1}]$. Given two  paths $p_1:\ [0, 1] \rightarrow\
\S$ and  $p_2:\ [0, 1] \rightarrow\ \S$ such that $p_1(1)=p_2(0)$,
we denote by $p_2\circ p_1$ the natural composition:
\begin{equation}
p_2\circ p_1(s) = \cases{p_1(2s),& for $s\in [0, {1\over 2}]$ \cr
       p_2(2s-1), &  for $s\in [{1\over 2}, 1]$.\cr}
\end{equation}
The ``inverse" of a path $p: [0, 1] \rightarrow \S$ is a path
\begin{equation}
p^{-1}(s) := p(1 - s).
\end{equation}
A path which starts and ends at the same point is called a {\it loop}.
We will be interested in {\it based} loops. Let us therefore fix, once
and for all, a point $\xz\in \S$. Denote by $\L$ the set of piecewise
$C^\w$ loops which are based at $\xz$, i.e., which start and end at
$\xz$.

Given a connection $A$ on $P(\S, SU(2))$, a loop $\a\in \L$, and a
fixed point ${\hat \xz}$ in the fiber over the point $\xz$, we denote
the corresponding element of the holonomy group by $H(\a, A)$. (Since
$P$ is a product bundle, we will make the obvious choice ${\hat \xz}=
(\xz, e)$, where $e$ is the identity in $SU(2)$.) Using the space of
connections $\A$, we now introduce a key equivalence relation on $\L$:

{\narrower\smallskip\noindent{\sl Two loops $\a, \b \in \L$ will be
said to be holonomically equivalent, $\a \sim \b$, iff $H(\a, A) =
H(\b, A)$, for {\rm every} SU(2)--connection $A$ in $\A$.} \smallskip}

\noindent
Each holonomic equivalence class will be called a {\it hoop}. Thus,
for example, two loops (which have the same orientation and) which
differ only in their parametrization define the same hoop. Similarly,
if two loops differ just by a retraced segment, they belong to the
same hoop.  (More precisely, if paths $p_1$ and $p_2$ are such that
$p_1(1) = p_2(0)$ and $p_1(0) = p_2(1) = \xz$, so that $p_2\circ p_1$
is a loop in $\L$, and if $\rho$ is a path starting at the point
$p_1(1) \equiv p_2(0)$, then $p_2\circ p_1$ and $p_2\circ
\rho\circ\rho^{-1}\circ p_1$ define the same hoop). Note that in
general, the equivalence relation depends on the choice of the gauge
group, whence the hoop should in fact be called a $SU(2)$--hoop.
However, since the group is fixed to be $SU(2)$ in the main body of
this paper, we drop this suffix. The hoop to which a loop $\a$ belongs
will be denoted by $\at$.

The collection of all hoops will be denoted by $\HG$; thus $\HG =
\L/\!\sim$.  Note that, because of the properties of the holonomy map,
the space of hoops, $\HG$, has a natural group structure: $\at
\cdot\bt = \widetilde{\a \circ
\b}$. We will call it the {\it hoop group}
\footnote{The hoop equivalence relation seems to have been introduced
independently by a number of authors in different contexts (e.g.\ by
Gambini and collaborators in their investigation of Yang--Mills theory
and by Ashtekar in the context of 2+1 gravity). Recently, the group
structure of $\HG$ has been systematically exploited and extended by
Di Bartolo, Gambini and Griego [9] to develop a new and potentially
powerful framework for gauge theories.}.

With this machinery at hand, we can now introduce the A--I \C*. We begin
by assigning to every $\at\in \HG$ a real--valued function $T_{\at}$ on
$\A/\G$, bounded between $-1$ and $1$:
\begin{equation}
T_{\at }(\tilde{A}):=\textstyle{1\over 2}\Tr\ H(\a, {A}),
\end{equation}
where $\At \in \A/\G$; $A$ is any connection in the gauge equivalence
class $\At$; $\a$ any loop in the hoop $\at$; and where the trace
is taken in the
fundamental representation of $SU(2)$. ($T_{\at}$ is called the Wilson
loop function in the physics literature.) Due to $SU(2)$ trace
identities, product of these functions can be expressed as sums:
\begin{equation}
T_{\at}T_{\bt} = \textstyle{1\over 2} (T_{\at\cdot\bt } +
T_{\at\cdot\bt^{-1}}),
\end{equation}
where, on the right side, we have used the composition of hoops in
$\HG$.  Therefore, the complex vector space spanned by the $T_{\at}$
functions is closed under the product; it has the structure of a
$\star$--algebra.  Denote it by $\HA$ and call it the {\it holonomy
algebra}. Since each $T_{\at}$ is a bounded continuous function on
$\A/\G$, $\HA$ is a subalgebra of the \C* of all complex--valued,
bounded, continuous functions thereon.  The completion $\HAbar$ of
$\HA$ (under the sup norm) has therefore the structure of a \C*. This
is the A--I \C*.

As in [2], one can make the structure of $\HAbar$ explicit by
constructing it, step by step, from the space of loops $\L$. Denote by
$\FL$ the free vector space over complexes generated by $\L$. Let $K$
be the subspace of $\FL$ defined as follows:
\begin{equation}
\sum_i a_i \a_i \in K \quad{\rm iff}\quad \sum_i a_i T_{\at_i}(\At) = 0,
\ \ \forall \At\in \A/\G .
\end{equation}
It is then easy to verify that $\HA = \FL/K$. (Note that the
$K$--equivalence relation subsumes the hoop equivalence.) Thus,
elements of $\HA$ can be represented either as $\sum a_i T_{\at_i}$ or
as $\sum a_i [\a_i]_K$.  The $\star$--operation, the product and the
norm on $\HA$ can be specified directly on $\FL/K$ as follows:
\begin{eqnarray}
 \sum_i (a_i \ [\a]_K )^\star &=& \sum_i \bar{a_i}\ [\a]_K, \nonumber
 \\ (\sum_i a_i \ [\a]_K )\cdot(\sum_j b_j \ [\b]_K ) &=& \sum_{ij}
a_i b_j ([\a_i\circ\b_j]_K + [\a_i\circ\b_j^{-1}]_K), \nonumber\\
||\ \sum_i a_i\ [\a_i]_K\  || &=& \sup_{\At\in \A/\G} |\sum_i a_i\ T_{{\at}_i}
(\At)|,
\end{eqnarray}
where $\bar{a}_i$ is the complex conjugate of $a_i$. The \C* $\HAbar$ is
obtained by completing $\HA$ in the norm--topology.

By construction, $\HAbar$ is an Abelian \C*. Furthermore, it is
equipped with the identity element: $T_{\tilde{o}} \equiv [o]_K$,
where $o$ is the trivial (i.e. zero) loop. Therefore, we can directly
apply the Gel'fand representation theory. Let us summarize the main
results that follow [2].  First, we know that $\HAbar$ is naturally
isomorphic to the $C^\star$ algebra of complex--valued, continuous
functions on a compact, Hausdorff space, the spectrum $sp(\HAbar)$.
Furthermore, one can show [5] that the space $\A/\G$ of connections
modulo gauge transformations is densely embedded in $sp(\HAbar)$.
Therefore, we can denote the spectrum as $\AGbar$. Second, every
continuous, cyclic representation of $\HAbar$ by bounded operators on
a Hilbert space has the following form: The representation space is
the Hilbert space $L^2(\AGbar, d\mu)$ for some regular measure $d\mu$
on $\AGbar$ and elements of $\HAbar$, regarded as functions on
$\AGbar$, act simply by (pointwise) multiplication. Finally, each of
these representations is uniquely determined (via the
Gel'fand--Naimark--Segal construction) by a continuous, positive
linear functional $<.>$ on the \C* $\HAbar: \sum a_i T_{\at_i}\mapsto
<\sum a_i T_{\at_i}> \equiv \sum a_i <T_{\at_i}>$.

The functionals $<.>$ naturally define functions $\Gamma(\a)$ on the
space $\L$ of loops --which in turn serve as the ``generating
functionals'' for the representation-- via: $\Gamma(\a) \equiv
<T_{\at}>$.  Thus, there is a canonical correspondence between regular
measures on $\AGbar$ and generating functionals $\Gamma(\a)$. The
question naturally arises: can we write down necessary and sufficient
conditions for a function on the loop space $\L$ to qualify as a
generating functional? Using the structure of the algebra $\HA$
outlined above, one can answer this question affirmatively:

{\narrower\smallskip\noindent\sl{ A function $\Gamma(\a)$ on $\L$
serves as a generating functional for the GNS construction if and only
if it satisfies the following two conditions:}
\smallskip}
\begin{displaymath}
\sum_i a_i\ T_{{\at}_i} = 0 \ \ \Rightarrow \sum_i a_i\ \Gamma (\a_i)= 0;
 \quad{\rm and}
\end{displaymath}
\begin{equation}
\sum_{i,j}\bar{a_i}a_j (\Gamma(\a_i\circ\a_j)
+ \Gamma(\a_i\circ\a_j^{-1})
\ge 0.
\end{equation}
\noindent
The first condition ensures that the functional $<.>$ on $\HA$,
obtained by extending $\Gamma(\a)$ by linearity, is well--defined
while the second condition (by (2.6)) ensures that it satisfies the
``positivity'' property, $<B^\star B>\ \ge 0, \ \forall B\in \HA$.
Thus, together, they provide a positive linear functional on the
$\star$-algebra $\HA$. Finally, since $\HA$ is dense in $\HAbar$ and
contains the identity, it follows that the positive linear functional
$<,>$ extends to the full $C^\star$-algebra $\HAbar$.

Thus, a loop function $\Gamma (\a)$ satisfying (2.9) determines a
cyclic representation of $\HAbar$ and therefore a regular measure on
$\AGbar$. Conversely, every regular measure on $\AGbar$ provides
(through vacuum expectation values) a loop functional $\Gamma(\a)
\equiv <T_{\at}>$ satisfying (2.9). Note, finally, that the first
equation in (2.9) ensures that the generating function factors through
the hoop equivalence relation: $\Gamma(\a) \equiv
\Gamma(\at)$.

We thus have an interesting interplay between connections and loops,
or, more precisely, generalized gauge equivalent connections (elements
of $\AGbar$) and hoops (elements of $\HG$). The generators $T_{\at}$
of the holonomy algebra (i.e., configuration operators) are labelled
by elements $\at$ of $\HG$. The elements of the representation space
(i.e. quantum states) are $L^2$ functions on $\AGbar$. Regular
measures $d\mu$ on $\AGbar$ are, in turn, determined by functions on
$\HG$ (satisfying (2.9)). The group of diffeomorphisms on $\S$ has a
natural action on the algebra $\HAbar$, its spectrum $\AGbar$, and the
space of hoops, $\HG$. The measure $d\mu$ is invariant under this
induced action on $\AGbar$ iff $\Gamma(\at)$ is invariant under the
induced action on $\HG$. Now, a diffeomorphism invariant function of
hoops is a function of generalized knots ---generalized because our
loops are allowed to have kinks, intersections and segments that
may be traced more than once.
\footnote{ If the gauge group were more general, we
could not have expressed products of the generators $T_{\at}$ as sums
(Eq.\ (2.6)). For $SU(3)$, for example, one can not get rid of double
products; triple (and higher) products can be however reduced to
linear combinations of double products of $T_{\at}$ and the $T_{\at}$
themselves. In this case, the generating functional is defined on
single {\it and double} loops, whence, in addition to knot invariants,
link invariants can also feature in the definition of the generating
function.}.
Hence, there is a 1--1 correspondence between generalized knot
invariants (which in addition satisfy (2.9)) and regular
diffeomorphism invariant measures on generalized gauge equivalent
connections.

\section {The spectrum}

The constructions discussed in Section 2 have several appealing
features.  In particular, they open up an algebraic approach to the
integration theory on the space of gauge equivalent connections and
bring to forefront the duality between loops and connections. However,
in practice, a good control on the precise content and structure of
the Gel'fand spectrum $\AGbar$ of $\HAbar$ is needed to make further
progress.  Even for simple algebras which arise in non--relativistic
quantum mechanics---such as the algebra of almost periodic functions
on $R^n$---the spectrum is rather complicated; while $R^n$ is densely
embedded in the spectrum, one does not have as much control as one
would like on the points that lie in its closure (see, e.g., [5]). In
the case of a \C* of all continuous, bounded functions on a completely
regular topological space, the spectrum is the Stone--C\v ech
compactification of that space, whence the situation is again rather
unruly. In the case of the A--I algebra $\HAbar$, the situation seems
to be even more complicated: Since the algebra $\HA$ is generated only
by {\it certain} functions on $\A/\G$, at least at first sight,
$\HAbar$ appears to be a {\it proper} sub--algebra of the \C* of all
continuous functions on $\A/\G$, and therefore outside the range of
applicability of standard theorems.

However, the holonomy \C* is also rather special: it is constructed
from natural geometrical objects---connections and loops---on the
3--manifold $\S$.  Therefore, one might hope that its spectrum can be
characterized through appropriate geometric constructions. We shall
see in this section that this is indeed the case. The specific
techniques we use rest heavily on the fact that the loops involved are
all piecewise {\it analytic}.

This section is divided into three parts. In the first, we introduce
the key tools that are needed (also in subsequent sections), in the
second, we present the main result and in the third we give examples
of ``generalized gauge equivalent connections'', i.e., of elements of
$\AGbar - \A/G$. To keep the discussion simple, generally we will not
explicitly distinguish between paths and loops and their images in
$\S$.

\subsection{Loop decomposition}
A key technique that we will use in various constructions is the
decomposition of a finite number of hoops, $\at_1, ... ,\at_k$, into a
finite number of {\it independent} hoops, $\bt_1, ... ,\bt_n$. The
main point here is that a given set of loops need not be
``holonomically independent'': every open segment in a given loop may
be shared by other loops in the given set, whence, for {\it any}
connection $A$ in $\A$, the holonomies around these loops could be
inter--related. However, using the fact that the loops are piecewise
analytic, we will be able to show that any finite set of loops {\it
can} be decomposed into a finite number of independent segments and
this in turn leads us to the desired independent hoops. The
availability of this decomposition will be used in the next
sub--section to obtain a characterization of elements of $\AGbar$ and
in Section 4 to define ``cylindrical'' functions on $\AGbar$.

Our aim then is to show that, given a finite number of hoops, $\at_1, ...,
\at_k$, (with $\at_i \not=$ identity in $\HG$ for any $i$),
there exists a finite set of loops, $\{\b_1,...,\b_n\}\subset \L$,
which satisfies the following properties:

\begin{enumerate}
\item If we denote by ${\HG}(\gt_1,...,\gt_m)$ the subgroup of the hoop
group $\HG$ generated by the hoops $\gt_1, ...,\gt_m$, then, we have
$${\HG}(\at_1, ...\at_k)\subset {\HG}(\bt_1, ...,\bt_n), \eqno(3.1)$$
where, as before, ${\bt}_i$ denotes the hoop to which $\b_i$ belongs.
\item Each of the new loops $\b_i$ contains an open segment (i.e.
an embedding of an interval) which is traced exactly once and which
has at most a finite number of intersections

intersects any of the remaining loops $\b_j$ (with $i\not= j$) at most
at a finite number of points.
\end{enumerate}

\noindent
The first condition makes the sense in which each hoop $\at_i$ can be
decomposed in terms of $\bt_j$ precise while the second condition
specifies the sense in which the new hoops $\bt_1, ...,\bt_n$ are
independent.

Let us begin by choosing a loop $\a_i\in \L$ in each hoop $\at_i$, for
all $i=1,...,k$ such that none of the $\at_i$ has a piece which is
immediately retraced (i.e., none of the loops contain a path of the
type $\rho \cdot \rho^{-1}$). Now, {\it since the loops are piecewise
analytic, any two which overlap do so either on a finite number of
finite intervals and/or intersect in a finite number of points.}
Ignore the isolated intersection points and mark the end points of all
the overlapping intervals (including the end points of self overlaps
of a loop). Let us call these marked points {\it vertices}. Next,
divide each loop $\a_i$ into paths which join consecutive end points.
Thus, each of these paths is piecewise--analytic, is part of at least
one of the loops, $\a_1,...,\a_k$, and, has a non--zero (parameter)
measure along any loop to which it belongs. Two distinct paths
intersect at a finite set of points.  Let us call these (oriented)
paths {\it edges}. Denote by $n$ the number of edges that result.  The
edges and vertices form a (piecewise analytically embedded) graph. By
construction, each loop in the list is contained in the graph and,
ignoring parametrization, can be expressed as a composition of a
finite number of its $n$ edges.  Finally, this decomposition is
``minimal'' in the sense that, if the initial set $\a_1, ..., \a_k$ of
loops is extended to include $p$ additional loops, $\a_{k+1},
...,\-\a_{k+p}$, each edge in the original decomposition of $k$ loops
is expressible as a product of the edges which feature in the
decomposition of the $k+p$ loops.

The next step is to convert this edge--decomposition of loops into a
decomposition in terms of elementary loops. This is easy to achieve.
Connect each vertex $v$ to the base point $\xz$ by oriented, piecewise
analytic curve $q(v)$ such that these curves overlap with any edge
$e_i$ {\it at most} at a finite number of isolated points. Consider
the closed curves $\b_i$, starting and ending at the base point $\xz$,
$$ \b_i := q (v_i^+) \circ e_i\circ (q(v_i^{-}))^{-1}, \eqno(3.2)$$
where $v_i^\pm$ are the end points of the edge $e_i$, and denote by
${\cal S}$ the set of these $n$ loops. Loops $\b_i$ are {\it not}
unique because of the freedom in choosing the curves $q(v)$.  However,
we will show that they provide us with the required set of independent
loops associated with the given set $\{\a_1, ...,\a_k\}$.

Let us first note that the decomposition satisfies certain conditions
which follow immediately from the properties of the segments noted
above:

\begin{enumerate}
\item ${\cal S}$ is a finite set and every $\b_i\in {\cal S}$ is a
non--trivial loop in the sense that the hoop $\bt_i$ it defines is not
the identity element of $\HG$;
\item every loop $\b_i$ contains an open interval which is traversed
exactly once and no finite segment of which is shared by any other
loop in ${\cal S}$;
\item every hoop $\at_1,...,\at_k$ in our initial set can be
expressed as a finite composition of hoops defined by elements of
${\cal S}$ and their inverses (where a loop $\b_i$ and its inverse
$\b_i^{-1}$ may occur more than once); and,
\item if the initial set $\a_1, ...,\a_k$ of loops is extended to
include $p$ additional loops, $\a_{k+1}, ..., \- \a_{k+p}$, and if
${\cal S}'$ is the set of $n'$ loops $\b'_1, ..., \b'_{n'}$
corresponding to $\a_1, ..., \a_{k+p}$, such that the paths $q'(v')$
agree with the paths $q(v)$ whenever $v=v'$, then each hoop $\bt_i$
is a finite product of the hoops $\bt_j'$.
\end{enumerate}

\noindent
These properties will play an important role throughout the paper. For
the moment we only note that we have achieved our goal: the property
2 above ensures that the set $\{\b_1, ..., \b_n\}$ of loops is
independent in the sense we specified in the point 2 (just below
Eq. (3.1)) and that 3 above implies that the hoop group generated by
$\a_1, ...\a_k$ is contained in the hoop group generated by $\b_1,
...\b_n$, i.e. that our decomposition satisfies (3.1).

We will conclude this sub--section by showing that the independence of
hoops $\{\bt_1,...,\- \bt_n\}$ has an interesting consequence which
will be used repeatedly:

\begin{lemma}
%{\narrower\smallskip\noindent\sl{
For every $(g_1, ..., g_n) \in [SU(2)]^n$, there exists a connection
$A_o \in \A$ such that:
  $$ H(\b_i , A_o) = g_i \forall i = 1, ..., n\eqno(3.3)$$
\end{lemma}
%\smallskip}

\begin{proof}
Fix a sequence of elements $(g_1, ....,g_n)$ of $SU(2)$, i.e., a
point in $[SU(2)]^n$. Next, for each loop $\b_i$, pick a connection
${A'}_{(i)}$ such that its support intersects the $i$--th segment
$s_i$ in a finite interval, and, if $j\not= i$, intersects the $j$--th
segment $s_j$ only on a set of measure zero (as measured along any
loop $\a_m$ containing that segment).
 Consider the holonomy of
$A'_{(i)}$ around $\b_i$. Let us suppose that the connection is such
that the holonomy $H({\b_i},{A'}_{(i)})$ is not the identity element
of $SU(2)$. Then, it is of the form: $$ H({\b_i},{A'}_{(i)}) =
\exp{w},\ \ w\in su(2), \ w\not= 0$$
Moreover, let the connection
be abelian; i.e. may be written as ${A'}_{(i)}\ =\ w{a'}_{(i)}$,
 ${a'}_{(i)}$ being
a real valued 1-form on $\Sigma$.
 Now, consider a connection
$${A}_{(i)} = t\ g^{-1}\cdot{A'}_{(i)}\ \cdot g,\ \ g\in SU(2),\ \
t\in {R}$$ Then, we have: $$H(\b_i, A_{(i)}) = \exp t(g^{-1}w g).$$
Hence, by choosing $g$ and $t$ appropriately, we can make $H(\b_i,
A_{(i)})$ coincide with any element of $SU(2)$, in particular $g_i$.
Because of the independence of the loops $\b_i$ noted above, we can
choose connections ${A}_{(i)}$ independently for every $\b_i$. Then,
the connection
$$A_o :={A}_{(1)} +...+{A}_{(n)} \eqno(3.4)$$
satisfies (3.3).
\end{proof}

\begin{enumerate}
\item The result (3.3) we just proved can be taken as a statement
of the independence of the $(SU(2))$-hoops $\bt_1, ... ,\bt_n$; it
captures the idea that the loops $\b_1, ...,\b_n$ we constructed above
are holonomically independent as far as $SU(2)$ connections are
concerned.  This notion of independence is, however, weaker than the
one contained in the statement 2 above. Indeed, that notion does not
refer to connections at all and is therefore not tied to any specific
gauge group. More precisely, we have the following. We just showed
that if loops are independent in the sense of 2, they are necessarily
independent in the sense of (3.3). The converse is not true. Let $\a$
and $\g$ be two loops in $\L$ which do not share any point other than
$\xz$ and which have no self--intersections or self--overlaps. Set
$\b_1 = \a$ and $\b_2 = \a\cdot \g$. Then, it is easy to check that
$\bt_i, \ i=1,2$ are independent in the sense of (3.3) although they
are obviously not independent in the sense of 2. Similarly, given
$\a$, we can set $\b =\a^2$. Then, $\{\b\}$ is independent in the
sense of (3.3) (since square--root is well--defined in $SU(2)$) but
not in the sense of 2. From now on, we will say that loops satisfying
2 are {\it strongly independent} and those satisfying (3.3) are {\it
weakly independent}. This definition extends naturally to hoops.
Although we will not need them explicitly, it is instructive to spell
out some algebraic consequences of these two notions.  Algebraically,
weak independence of hoops $\bt_i$ ensures that $\bt_i$ freely
generate a sub-group of $\HG$. On the other hand, if $\bt_i$ are
strongly independent, then every homomorphism from
$\HG(\bt_1,...\bt_n)$ to {\it any} group $G$ can be extended to every
finitely generated subgroup of $\HG$ which contains $\HG(\bt_1, ...,
\bt_n)$. Weak independence will suffice for all considerations in this
section. However, to ensure that the measure defined in Section 4 is
well--defined, we will need hoops which are strongly independent.
\item The availability of the loop decomposition sheds considerable light
on the hoop equivalence: One can show that two loops in $\L$ define
the same $(SU(2))$-hoop if and only if they differ by a combination of
reparametrization and (immediate) retracings. This second
characterization is useful because it involves only the loops; no
mention is made of connections and holonomies.  This characterization
continues to hold if the gauge group is replaced by $SU(N)$ or indeed
any group which has the property that, for every non-negative integer
$n$, the group contains a subgroup which is freely generated by $n$
elements.
\end{enumerate}

\subsection {Characterization of $\AGbar$}
We will now use the availability of this hoop decomposition to obtain
a complete characterization of all elements of the spectrum $\AGbar$.

The characterization is based on results due to Giles [10]. Recall
first that elements of the Gel'fand spectrum $\AGbar$ are in 1--1
correspondence with (multiplicative, $\star$--preserving)
homomorphisms from $\HAbar$ to the $\star$--algebra of complex
numbers. Thus, in particular, every $\Ab\in \AGbar$ acts on
$T_{\at}\in \HAbar$ and hence on any hoop $\at$ to produce a complex
number $\Ab (\at)$. Now, using certain constructions due to Giles,
A--I [2] were able to show that:

\begin{lemma}
Every element $\Ab$ of $\AGbar$ defines a homomorphism
$\hat{H}_{\bar A}$ from the hoop group $\HG$ to $SU(2)$ such that,
$\forall \at\in\HG$,
$$\Ab (\at) = \textstyle {1\over 2}\ \Tr \ \hat{H}_{\bar{A}} (\at).
\eqno(3.5a)$$
\end{lemma}

They also exhibited the homomorphism explicitly. (Here, we have
paraphrased the A--I result somewhat because they did not use the
notion of hoops. However, the step involved is completely
straightforward.)

Our aim now is to establish the converse of this result.  Let us make
a small digression to illustrate why the converse can not be entirely
straightforward. Given a homomorphism $\hat{H}$ from $\HG$ to $SU(2)$
we can simply define $\Ab$ via: $\Ab(\at) = \textstyle{1\over 2}\Tr\
H(\at)$. The question is whether this $\Ab$ can qualify as an element
of the spectrum. From the definition, it is clear that:
$$|\Ab(\at)|\ \leq \ \ \|T_{\at} \| \equiv\sup_{\At\in \A/\G}
|T_{\at}(\At)|\eqno(3.6a)$$
for all hoops $\at$. On the other hand, to qualify as an element of
the spectrum $\AGbar$, the homomorphism $\Ab$ must be continuous,
i.e., should satisfy:
$$|\Ab(f)|\leq \|f\|\eqno(3.6b)$$
for every $f (\equiv \sum a_i [{\at}_i]_K) \in \HA$. Since, a priori,
(3.6b) appears to be stronger requirement than (3.6a), one would
expect that there are more homomorphisms $\hat{H}$ from the hoop group
$\HG$ to $SU(2)$ than the $\hat{H}_{\Ab}$, which arise from elements
of $\AGbar$. That is, one might expect that the converse of the A--I
result would not be true. It turns out, however, that if the holonomy
algebra is constructed from piecewise {\it analytic} loops, the
apparently weaker condition (3.6a) is in fact equivalent to (3.6b),
and the converse does hold. Whether it would continue to hold if one
used piecewise smooth loops---as was the case in the A--I
analysis---is not clear (see Appendix A).

We begin our detailed analysis with an immediate application of Lemma
3.1:

\begin{lemma}
For every homomorphism $\hat{H}$ from $\HG$ to $SU(2)$, and every
finite set of hoops $\{\at_1,...\at_k\}$ there exists an $SU(2)$
connection $A_o$ such that for every $\at_i$ in the set,
$$\hat{H}(\at_i) = H(\a_i,A_o),\eqno(3.7)$$ where, as before,
$H(\a_i,A_o)$ is the holonomy of the connection $A_o$ around any loop
$\a$ in the hoop class $\at$.
\end{lemma}

\begin{proof}
Let $(\b_1, ...,\b_n)$ be, as in Section 3.1, a set of independent
loops corresponding to the given set of hoops.  Denote by
$(g_1,...,g_n)$ the image of $(\bt_1,...,\bt_n)$ under $\hat{H}$. Use
the construction of Section 3.1 to obtain a connection $A_o$ such that
$g_i = H(\bt_i, A_o)$ for all $i$. It then follows (from the
definition of the hoop group) that for {\it any} hoop $\gt$ in the
sub--group $\HG{}(\bt_1, ...\bt_n)$ generated by $\bt_1, ...,\bt_n$,
we have $\hat{H}(\gt) = H(\gt, A_o)$. Since each of the given $\at_i$
belongs to $\HG{}(\bt_1, ...,\bt_n)$, we have, in particular, (3.7).
\end{proof}

We now use this lemma to prove the main result. Recall that each
$\Ab\in\AGbar$ is a homomorphism from $\HAbar$ to complex numbers and
as before denote by $\Ab(\at)$ the number assigned to $T_{\at} \in
\HAbar$ by $\Ab$. Then, we have:

\begin{lemma}
Given a homomorphism $\hat{H}$ from the hoop group $\HG$ to $SU(2)$
there exists an element ${\Ab}_{\hat H}$ of the spectrum $\AGbar$ such
that
$$\Ab_{\hat H} (\at) = \textstyle{1\over 2}\Tr\ \hat{H}(\at).\eqno(3.5b)$$
Two homomorphisms $\hat{H}$ and $\hat{H}'$ define the same element of
the spectrum if and only if $\hat{H}'(\at ) = g^{-1}\cdot \hat{H}(\at)
\cdot g,\ \ \forall \at\in \HG$ for some ($\at$--independent) $g$ in $SU(2)$.
\end{lemma}

\begin{proof}
The idea is to {\it define} the required $\Ab_{\hat H}$ using (3.5b),
i.e. to show that the right side of (3.5b) provides a
homomorphism from $\HAbar$ to the $\star$--algebra of complex numbers.
Let us begin with the free complex vector space ${\cal F}\HG$ over the
hoop group $\HG$ and define on it a complex--valued function $h$ as
follows: $h(\sum a_i {\at}_i) := \textstyle{1\over 2} \sum a_i\ \Tr\
\hat{H}(\at_i)$. We will first show that $h$ passes through the
$K$--equivalence relation of A--I (noted in Section 2) and is
therefore a well--defined complex--valued function on the holonomy
$\star$--algebra $\HA = {\cal FHG}/K$. Note first that, Lemma 3.3
immediately implies that, given a finite set of hoops, $\{{\at}_1,
..., {\at}_k\}$, there exists a connection $A_o$ such that,
 $$|h(\sum_{i=1}^k a_i {\at}_i)| = |\sum_{i=1}^k \ a_i
   T_{\at_i}(\At_o) | \le \ \sup_{\At\in \A/\G} |\sum_{i=1}^k\
   a_iT_{{\at}_i}(\At)|.\eqno(3.8)$$
Therefore, we have:
$$\sum_{i=1}^k a_i T_{{\at}_i}(\At ) = 0 \ \forall \At\in \AG
\quad\Rightarrow\quad h(\sum_{i=1}^k a_i {\at}_i) = 0. $$
Since the left side of this implication defines the $K$--equivalence,
it follows that the function $h$ on ${\cal FHG}$ has a well--defined
projection to the holonomy $\star$--algebra $\HA$. That $h$ is linear
is obvious from its definition. That it respects the
$\star$--operation and is a multiplicative homomorphism follows from
the definitions (2.8) of these operations on $\HA$ (and the definition
of the hoop group $\HG$). Finally, the continuity of this funtion on
$\HA$ is obvious from (3.8). Hence $h$ extends uniquely to a
homomorphism from the \C* $\HAbar$ to the $\star$--algebra
of complex numbers, i.e. defines a unique element ${\At}_{\hat{H}}$
via ${\At}_{\hat{H}}(B) = h(B),\ \forall B\in \HAbar$.

Next, suppose $\hat{H}_1$ and $\hat{H}_2$ give rise to the same
element of the spectrum. In particular, then, $\Tr\ \hat{H}_1(\at ) =
\Tr\ \hat{H}_2 (\at)$ for all hoops. Now, there is a general result:
Given two homomorphisms $\hat{H}_1$ and $\hat{H}_2$ from a group ${\bf
G}$ to $SU(2)$ such that $\Tr\ \hat{H}_1({\bf g}) = \Tr\
\hat{H}_2({\bf g}), \ \forall {\bf g}\in {\bf G}$, there exists
$g_o\in SU(2)$ such that $\hat{H}_2({\bf g}) = g_o^{-1}\cdot
\hat{H}_1({\bf g})\cdot g_o$, where $g_o$ is independent of ${\bf g}$.
Using the hoop group $\HG$ for ${\bf G}$, we obtain the desired
uniqueness result.
\end{proof}

To summarize, by combining the results of the three lemmas, we obtain
the following characterization of the points of the Gel'fand spectrum
of the holonomy \C*:

\begin{theorem}  Every point $\Ab$ of $\AGbar$ gives rise to a homomorphism
$\hat{H}$ from the hoop group $\HG$ to $SU(2)$ and every homomorphism
$\hat{H}$ defines a point of $\AGbar$, such that $\Ab (\at) =
\textstyle{1\over 2}\ \Tr\ \hat{H}(\at)$. This is a 1--1
correspondence modulo the trivial ambiguity that homomorphisms
$\hat{H}$ and $g^{-1}\cdot\hat{H}\cdot g$ define the same point of
$\AGbar$.
\end{theorem}

\noindent We conclude this sub--section with some remarks:

\begin{enumerate}
\item It is striking  that the Theorem 3.5 {\it does not require
the homomorphism} $\hat{H}$ {\it to be continuous}; indeed, no
reference is made to the topology of the hoop group anywhere. This
purely algebraic characterization of the elements of $\AGbar$ makes it
convenient to use it in practice.
\item Note that the homomorphism $\hat{H}$ determines an element of
$\AGbar$ and {\it not} of $\AG$; ${\Ab}_{\hat{H}}$ is {\it not}, in
general, a smooth (gauge equivalent) connection. Nonetheless, as Lemma
3.3 tells us, it can be approximated by smooth connections in the
sense that, given any {\it finite} number of hoops, one can construct
a smooth connection which is indistinguishable from $\Ab_{\hat{H}}$ as
far as these hoops are concerned. (This is stronger than the statement
that $\AG$ is dense in $\AGbar$.) Necessary and sufficient conditions
for $\hat{H}$ to arise from a smooth connection were given by Barrett
[11] (see also [7]).
\item There are several folk theorems in the literature to the
effect that given a function on the loop space $\L$ satisfying certain
conditions, one can reconstruct a connection (modulo gauge) such that
the given function is the trace of the holonomy of that connection.
(For a summary and references, see, e.g., [4].) Results obtained in
this sub--section have a similar flavor. However, there is a {\it key}
difference: Our main result shows the existence of a {\it generalized}
connection (modulo gauge), i.e., an element of $\AGbar$ rather than a
regular connection in $\AG$. A generalized connection can also be
given a geometrical meaning in terms of parallel transport, but in a
{\it generalized} bundle [7].
\end{enumerate}

\subsection{Examples of $\Ab$}
Fix a connection $A\in \A$. Then, the holonomy $H(\a, A)$ defines a
homomorphism $\hat{H}_A: {\HG} {\mapsto SU(2)}$. A gauge equivalent
connection, $A' = g^{-1}Ag +g^{-1}dg$, gives rise to the homomorphism
$\hat{H}_{A'}= g^{-1}(\xz)\hat{H}_{A}\ g(\xz)$. Therefore, by Theorem
3.5, $A$ and $A'$ define the same element of the Gel'fand spectrum;
$\A/\G$ is naturally embedded in $\AGbar$. Furthermore, Lemma 3.3 now
implies that the embedding is in fact dense. This provides an
alternate proof of the A--I and Rendall%
\footnote{
Note, however, that while this proof is tailored to the holonomy \C*
$\HAbar$, Rendall's [5] proof is applicable in more general contexts.}
results quoted in Section 1.  Had the gauge group been different
and/or had $\S$ a dimension greater than 3, there would exist
non--trivial $G$--principal bundles over $\S$. In this case, even if
we begin with a specific bundle in our construction of the holonomy
\C*, connections on {\it all} possible bundles belong to the Gel'fand
spectrum (see Appendix B). This is an one illustration of the
non--triviality of the
completion procedure leading from $\A/\G$ to $\AGbar$.

In this subsection, we will illustrate another facet of this
non--triviality: We will give examples of elements of $\AGbar$ which
do not arise from $\A/\G$.  These are the {\it generalized} gauge
equivalent connections $\Ab$ which have a ``distributional character''
in the sense that their support is restricted to sets of measure zero
in $\S$.

Recall first that the holonomy of the zero connection around any hoop
is identity. Hence, the homomorphism $\hat{H}_o$ from $\HG$ to $SU(2)$
it defines sends every hoop $\at$ to the identity element $e$ of
$SU(2)$ and the multiplicative homomorphism it defines from the \C*
$\HAbar$ to complex numbers sends each $T_{\at}$ to 1 ($=(1/2) \Tr\
e)$. We now use this information to introduce the notion of the support
of a generalized connection. We will say that $\Ab \in
\AGbar$ has support on a subset $S$ of $\S$ if for every hoop $\at$
which has at least one representative loop $\a$ which fails to pass
through $S$, we have: i) $\hat{H}_{\Ab}(\at) = e$ , the identity
element of $SU(2)$; or, equivalently, ii) $\Ab ([\at]_K) = 1$. (In
ii), $\Ab$ is regarded as a homomorphism of $\star$--algebras.
While $\hat{H}_{\Ab}$ has an ambiguity, noted in Theorem 3.5,
conditions i) and ii) are insensitive to it.)

We are now ready to give the simplest example of a generalized
connection $\Ab$ which has support at a single point $x \in \S$. Note
first that, due to piecewise analyticity, if $\a \in \L$ passes
through $x$, it does so only a finite number of times, say $N$. Let us
denote the incoming and outgoing tangent directions to the curve at
its $j$th passage through $x$ by $(v^{-}_j, v^{+}_j)$.  The
generalized connections we want to now define will depend only on
these tangent directions at $x$. Let $\phi$ be a (not necessarily
continuous) mapping from the 2--sphere $S^2$ of directions in the
tangent space at $x$ to $SU(2)$. Set:
$$
\Phi_j = \phi(-v^{-}_j)^{-1}\cdot \phi(v^{+}_j),
$$
so that, if at the $j$--th passage, $\a$ arrives at $x$ and then
simply returns by retracing its path in a neighborhood of $x$, we have
$\Phi_j = e$, the identity element of $SU(2)$. Now, define
$\hat{H}_\phi: {\HG} {\mapsto SU(2)}$ via: $$ \hat{H}_\phi (\at) :=
\cases{e, & if $\at\notin {\HG}_x$,\cr \Phi_N\ \ ...\ \ \Phi_1, &
otherwise\cr}\eqno(3.9)$$ where ${\HG}_x$ is the space of hoops every
loop in which passes through $x$. For each choice of $\phi$, the
mapping $\hat{H}_\phi$ is well--defined, i.e., is independent of the
choice of the loop $\a$ in the hoop class $\at$ used in its
construction. (In particular, we used tangent directions rather than
vectors to ensure this.) It defines a homomorphism from $\HG$ to
$SU(2)$ which is non--trivial only on ${\HG}{}_x$, and hence a
generalized connection with support at $x$.  Furthermore, one can
verify that (3.9) is the most general element of $\AGbar$ which has
support at $x$ and depends only on the tangent vectors at that point.
Using analyticity of the loops, we can similarly construct elements
which depend on the higher order behavior of loops at $x$.(There is no
generalized connection with support at $x$ which depends only on the
zeroth order behavior of the curve, i.e., only on whether the curve
passes through $x$ or not.)

One can proceed in an analogous manner to produce generalized
connections which have support on $1$ or $2$--dimensional
sub--manifolds of $\S$. We conclude with an example. Fix a
$2$--dimensional, oriented, analytic sub--manifold $M$ of $\S$ and an
element $g$ of $SU(2)$, ($g\not= e$). Denote by ${\HG}{}_M$ the subset
of $\HG$ consisting of hoops, each loop of which intersects $M$
(non--tangentially, i.e., such that least one of the incoming or the
outgoing tangent direction to the curve is transverse to $M$.)  As
before, due to analyticity, any $\a\in\L$ can intersect $M$ only a
finite number of times. Denote as before the incoming and the outgoing
tangent directions at the $j$--th intersection by $v^{-}_j$ and
$v^{+}_j$ respectively.  Let $\epsilon(\a^\pm_j)$ equal $1$ if the
orientation of $(v^\pm_j ,M)$ coincides with the orientation of $\S$;
$-1$ if the two orientations are opposite; and $0$ if $v^\pm_j$ is
tangential to $M$.  Define $\hat{H}_{(g,M)}: {\HG}{\mapsto SU(2)}$
via:
$$ \hat{H}_{(g, M)}(\at) := \cases{e, & if $\at\notin
{\HG}_M$,\cr g^{\epsilon(\a^{+}_N)} \cdot g^{\epsilon(\a^{-}_N)}\ \ \
... \ \ \ g^{\epsilon(\a^{+}_1)}\cdot g^{\epsilon(\a^{-}_1)}, &
otherwise, \cr}
\eqno(3.10)$$
where $g^0\equiv e$, the identity element of $SU(2)$. Again, one can
check that (3.10) is a well--defined homomorphism from $\HG$ to
$SU(2)$, which is non--trivial only on ${\HG}_M$ and therefore defines
a generalized connection with support on $M$. One can construct more
sophisticated examples in which the homomorphism is sensitive to the
higher order behavior of the loop at the points of intersection and/or
the fixed $SU(2)$ element $g$ is replaced by $g(x): M\mapsto SU(2)$.

\section {Integration on $\AGbar$}
In this section we shall develop a general strategy to perform
integrals on $\AGbar$ and introduce a natural, faithful,
diffeomorphism invariant measure on this space. The existence of this
measure provides considerable confidence in the the ideas involving
the loop--transform and the loop representation introduced in [3,4].

The basic strategy is to carry out an appropriate generalization of the
theory of cylindrical measures [12,13] on topological vector spaces
\footnote{This idea was developed also by Baez [6] using, however,
an approach which is somewhat different from the one presented here.
Chronologically, the authors of this paper first found the faithful
measure introduced in this section and reported this result in the
first draft of this paper. Subsequently, Baez and the authors
independently realized that the theory of cylindrical measures
provides the appropriate conceptual setting for this discussion.}.
The key idea in our approach is to replace the standard linear duality
on vector spaces by the duality between connections and loops. In the
first part of this section, we will define cylindrical functions and
in the second part, we will present a natural cylindrical measure.

\subsection{Cylindrical functions}
Let us begin by recalling the situation in the linear case.  Let $V$
denote a topological vector space and $V^\star$ its dual. Given a {\it
finite dimensional} subspace $S^\star$ of $V^\star$, we can define an
equivalence relation on $V$: $v_1 \sim v_2$ if and only if $<v_1 ,
s^*> = <v_2 , s^*>$ for all $v_1,v_2 \in V$ and $s^*\in S^\star$,
where $<v,s^*>$ is the action of $s^*\in S^\star$ on $v\in V$. Denote
the quotient $V/\!\sim$ by $S$. Clearly, $S$ is a finite--dimensional
vector space and we have a natural projection map $\pi(S^\star):
V\mapsto S$.  A function $f$ on $V$ is said to be {\it cylindrical} if
it is a pull--back via $\pi(S^\star)$ to $V$ of a function $\tilde{f}$
on $S$, for {\it some} choice of $S^\star$. These are the functions we
will be able to integrate.  Equip each finite--dimensional vector
space $S$ with a measure $d\tilde\mu$ and set $$\int_V\ d\mu\ f :=
\int_S\ d\tilde\mu \ \tilde{f}. \eqno(4.1)$$ For this definition to be
meaningful, however, the set of measures $d\tilde{\mu}$ that we chose
on vector spaces $S$ must satisfy certain compatibility conditions.
Thus, if a function $f$ on $V$ is expressible as a pull--back via
$\pi(S^\star_j)$ of functions $\tilde{f}_j$ on $S_j$, for $j = 1,2$,
say, we must have
$$\int_{S_1}\ d\tilde\mu_1 \ \tilde{f}_1 = \int_{S_2}\ d\tilde\mu_2 \
\tilde{f}_2.
\eqno(4.2)$$
Such compatible measures do exist. Perhaps the most familiar example
is the (normalized) Gaussian measure.

We want to extend these ideas, using $\AGbar$ in place of $V$. An
immediate problem is that $\AGbar$ is a genuinely non--linear space
whence there is no obvious analog of $V^\star$ and $S^\star$ which are
central to the above construction of cylindrical measures. The idea is
to use the results of Section 3 to find suitable substitutes for these
spaces. The appropriate choices turn out to be the following: we will
let the hoop group $\HG$ to play the role of $V^\star$ and its
subgroups, generated by a finite number of independent hoops, the role
of $S^\star$. (Recall that, $S^\star$ is generated by a finite number
of linearly independent basis vectors.)

More precisely, we proceed as follows. From Theorem 3.5, we know that
each $\Ab$ in $\AGbar$ is completely characterized by the homomorphism
$\hat{H}_{\Ab}$ from the hoop group $\HG$ to $SU(2)$, which is unique
up to the freedom $\hat{H}_{\Ab} \mapsto g_o^{-1} \hat{H}_{\Ab}\ g_o$
for some $g_o\in SU(2)$. Now suppose we are given $n$ loops, $\b_1,
..., \b_n$, which are (strongly) independent in the sense of Section
3.  Denote by $\Sstar$ the sub--group of the hoop group $\HG$ they
generate. Using this $\Sstar$, let us introduce the following
equivalence relation on $\AGbar$: ${\Ab}_1 \sim\Ab_2$ iff
$\hat{H}_{{\Ab}_1}(\gt) = g_o^{-1}\hat{H}_{{\Ab_2}}(\gt)\ g_o$ for all
$\gt\in \Sstar$ and some (hoop independent) $g_o\in Su(2)$. Denote by
$\pi(\Sstar )$ the projection from $\AGbar$ onto the quotient space
$[\AGbar ]/\!\sim$. The idea is to consider functions $f$ on $\AGbar$
which are pull--backs under $\pi(\Sstar)$ of functions $\tilde{f}$ on
$[\AGbar]/\!\sim$ and define the integral of $f$ on $\AGbar$ to be
equal to the integral of $\tilde{f}$ on $[\AGbar]/\!\sim$.  However,
for this strategy to work, $[\AGbar ]/\!\sim$ should have a simple
structure that one can control. Fortunately, this is the case:

\begin{lemma}
$[\AGbar]/\!\sim$ is isomorphic with $[SU(2)]^n/{\rm Ad}$.
\end{lemma}

\begin{proof}
By definition of the equivalence relation $\sim$, every $\{\Ab\}$ in
$[\AGbar] /\!\sim$ is in 1--1 correspondence with the restriction to
$\Sstar$ of an equivalence class $\{\hat{H}_{\Ab}\}$ of homomorphisms
from the full hoop group $\HG$ to $SU(2)$. Now, it follows from Lemma
3.3 that every homomorphism from $\Sstar$ to $SU(2)$ extends to a
homomorphism from the full hoop group to $SU(2)$. Therefore, there is
a 1--1 correspondence between equivalence classes $\{ A\}$ and
equivalence class of homomorphisms from $\Sstar$ to $SU(2)$ (where, as
usual, two are equivalent if they differ by the adjoint action of
$SU(2)$). Next, since $\Sstar$ is generated by the $n$ hoops $\bt_1,
..., \bt_n$, every homomorphism from $\Sstar$ to $SU(2)$ is completely
determined by its action on the $n$ generators. Hence, $\{\Ab\}$ is in
1--1 correspondence with the equivalence class of $n$--tuples,
$\{\hat{H}_{\Ab}(\bt_1), ..., \hat{H}_{\Ab}(\bt_n)\}
\equiv  \{g_o^{-1}\hat{H}_{\Ab}(\bt_1)\ g_o , ..., g_o^{-1}\hat{H}_{\Ab}
(\bt_n)\ g_o \}$, which in turn defines a point of $[SU(2)]^n/{\rm
Ad}$.  Thus, we have an injective map from $[\AGbar]/\!\sim$ to
$[SU(2)]^n/{\rm Ad}$.  Finally, it follows from Lemma 3.3 that the map
is also surjective. Thus $[\AGbar]/\!\sim$ is isomorphic with
$[SU(2)]^n/{\rm Ad}$.
\end{proof}

\noindent Some remarks about this result are in order:

\begin{enumerate}
\item Although we began with (strongly) independent loops $\{\b_1, ...,
\b_n\}$, the equivalence relation we introduced makes direct reference only
to the subgroup $\Sstar$ of the hoop group they generate. This is
similar to the situation in the linear case where one may begin with a
set of linearly independent elements $s_1^*, ..., s_n^*$, let them
generate a subspace $S^\star$ and then introduce the equivalence
relation on $S$ using $S^\star$ directly. There is, however, a
difference. In the linear case, $S^\star$ admits other bases and we
should make sure that our subsequent constructions refer only to
$S^\star$ and not to any specific basis therein.  In the case under
consideration, the situation is simpler: it follows from the notion of
(strong) independence of loops that the {\it only} set of independent
hoops which generates (precisely) $\Sstar$ is the set $\{\bt_1, ...,
\bt_n\}$ we began with. %CHECK!!
\item An important consequence of this remark is that, given
the sub--group $\Sstar$ of the hoop group $\HG$, the isomorphism
between $[\AGbar]/\!\sim$ and $[SU(2)]^n/{\rm Ad}$ is natural. That
is, although we have used the independent generators $\bt_1,
...,\bt_n$ in Lemma 4.1, since these generators are canonical, we do
not have to worry about how the identification between points of
$[\AGbar]/\!\sim$ and $[SU(2)]^n/{\rm Ad}$ would change if we change
the generators. Consequently, the projection map $\pi(\Sstar)$ can now
be taken to map $\AGbar$ to $[SU(2)]^n/{\rm Ad}$. In the case of
topological vector spaces, by contrast, one does have to worry about
this issue. In that case, it is only when a basis is specified in
$S^\star$ that we can identify points in $V/\!\sim$ with vectors in
$R^n$; if the basis is rotated, the identification changes
accordingly. Consequently, the projection map from $V$ to
$n$--dimensional vector spaces also depends on the choice of basis.
Although the space of cylindrical functions is unaffected by this
ambiguity, care {\it is} needed in, e.g., the specification of the
measure. Specifically, one must make sure that the values of integrals
of cylindrical functions are basis--independent.
\item The equivalence relation $\sim$ of Lemma 4.1
says that two generalized connections are equivalent if their actions
on the finitely generated subgroup $\Sstar$ of the hoop group are
indistinguishable. It follows from Lemma 3.3 that each equivalence
class $\{\Ab\}$ contains a regular connection $A$. Finally, if $A$
and $A'$ are two regular connections whose pull--backs to all $\b_i$
are the same for $i = 1,...,n$, then they are equivalent (where $\b_i$
is any loop in the hoop class $\bt_i$). Thus, in effect, the
projection $\pi(\Sstar)$ maps the domain space $\AGbar$ of the
continuum quantum theory to the domain space of a lattice gauge theory
where the lattice is formed by the $n$ loops $\b_1, ... ,\b_n$. The
initial choice of the independent hoops is, however, arbitrary and it
is this arbitrariness that is responsible for the richness of the
continuum theory.
\end{enumerate}
\medskip

We can now define cylindrical functions on $\AGbar$. As remarked
already, one can regard $\pi(\Sstar)$ as a projection map from
$\AGbar$ onto $[SU(2)]^n/{\rm Ad}$. By a {\it cylindrical function}
$f$ on $\AGbar$, we shall mean the pull--back to $\AGbar$ of a
continuous function $\tilde{f}$ on $(SU(2))^n/{\rm Ad}$ under
$\pi({\cal S}^\star)$ for {\it some} finitely generated sub--group
${\cal S}^\star$ of the hoop group. These are the functions we want to
integrate. Let us note an elementary property which will be used
repeatedly. Let $\Sstar$ be generated by $n$ strongly independent
hoops $\bt_1, ... ,\bt_n$ and $(\Sstar)'$ by $n'$ strongly independent
hoops $\bt'_1,...,\bt'_{n'}$ such that $(\Sstar)' \subseteq (\Sstar)$.
(Thus, $n'\le n$.) Then, it is easy to check that if a function $f$ on
$\AGbar$ is cylindrical with respect to $(\Sstar)'$, it is also
cylindrical with respect to $\Sstar$. Furthermore, there is now a
natural projection from $[SU(2)]^{n}$ on to $[SU(2)]^{n'}$ and
$\tilde{f}$ is the pull--back of $(\tilde{f})'$ via this projection.

In the remainder of this sub--section, we will analyse the structure
of the space $\c$ of cylindrical functions. First we have:

\begin{lemma}
$\c$ has the structure of a $\star$--algebra.
\end{lemma}

\begin{proof}
It is clear by definition of $\c$ that if $f\in \c$, then any complex
multiple $\lambda f$ as well as the complex--conjugate $\bar{f}$ also
belongs to $\c$. Next, given two elements, $f_i$ with $i=1,2$, of
$\c$, we will show that their sum and their product also belong to
$\c$. Let $\Sstar_i$ be the finitely generated subgroups of the hoop
group $\HG$ such that $f_i$ are the pull--backs to $\AGbar$ of
continuous functions $\tilde{f}_i$ on $[SU(2)]^{n_i}/ {\rm Ad}$.
Consider the set of $n_1+n_2$ independent hoops generating $\Sstar_1$
and $\Sstar_2$. While the first $n_1$ and the last $n_2$ hoops in this
set are independent, there may be relations between hoops belonging to
the first set and those belonging to the second. Using the technique
of Section 3.1, find the set of independent hoops, say,
$\bt_1,...,\bt_n$ which generate the given $n_1+n_2$ hoops and denote
by $\Sstar$ the subgroup of the hoop group generated by $\bt_1,
...,\bt_n$. Then, since $\Sstar_i\subseteq \Sstar$, it follows that
$f_i$ are cylindrical also with respect to $\Sstar$. Therefore,
$f_1+f_2$ and $f_1\cdot f_2$ are also cylindrical with respect to
$\Sstar$. Thus $\c$ has the structure of a $\star$--algebra.

\end{proof}

 Now,
since $[SU(2)]^n/{\rm Ad}$ is compact for any $n$, and since elements
of $\c$ are pull--backs to $\AGbar$ of all continuous functions on
$[SU(2)]^n/{\rm Ad}$, it follows that one can use the $sup$--norm to
endow $\c$ the structure of a \C*.

A natural class of functions one would like to integrate on $\AGbar$
is given by the (Gel'fand transforms of the) traces of holonomies. The
question then is if these functions are contained in $\c$. Not only is
the answer affirmative, but in fact the traces of holonomies generate
all of $\c$.  More precisely, we have:

\begin{theorem}
The completion $\cbar$ of $\c$ is isomorphic to the \C* $\HAbar$.
\end{theorem}

\begin{proof}
Let us begin by showing that $\HAbar$ is embedded in $\cbar$.  Given an
element $F$ with $F(A)= \sum a_i T_{{\at}_i}(A)$ of $\HAbar$, its
Gel'fand transform is a function $f$ on the spectrum $\AGbar$, given
by $f(\Ab) =\sum a_i \Ab(\at_i)$. Consider the set of hoops
$\at_1,...,\at_k$ that feature in the sum, decompose them into
independent hoops $\bt_1, ...,\bt_n$ as in Section 3.1 and denote by
$\Sstar$ the subgroup of the hoop group they generate. Then, it is
clear that the function $f$ is the pull--back via $\pi(\Sstar)$ of a
continuous function $\tilde{f}$ on $[SU(2)]^n/{\rm Ad}$. Hence we have
$f\in \c$.  Finally, since the norms of the functions $f$ and
$\tilde{f}$ are given by
  $$||f|| = \sup_{\Ab\in\AGbar} \ |f(\Ab)| \quad {\rm and} \quad
   ||\tilde{f}|| = \sup_{\bar{a}\in [SU(2)]^n/{\rm ad}}\ \
   |\tilde{f}(\bar{a})|,$$
it is clear that the map from $f$ to $\tilde{f}$ is isometric. The
first of these norms features in the definition of the (Gel'fand
transform of the)\C* $\HAbar$ while the second features in the \C*
$\cbar$. Finally, since $\HAbar$ is obtained by a $C^\star$--completion
of the space of functions of the form $F$, we have the result that
$\HAbar$ is embedded in the \C* $\cbar$.

Next, we will show that there is also an inclusion in the opposite
direction.  We first note a fact about $SU(2)$. Fix $n$--elements
$(g_1, ...,g_n)$ of $SU(2)$. Then, from the knowledge of traces (in
the fundamental representation) of all elements which belong to the
group generated by these $g_i, i=1,...,n$, we can reconstruct $(g_1,
..., g_n)$ modulo an adjoint map. It follows from this fact that the
space of functions $\tilde{f}$ on $[SU(2)]^n/{\rm Ad}$ which is the
image under the projection $\pi(\Sstar)$ of $\HAbar$ suffices to
separate points of $[SU(2)]^n/{\rm Ad}$. Since this space is compact,
it follows (from the Weirstrass theorem) that the \C* generated by
these projected functions is the entire \C* of continuous functions on
$[SU(2)]^n/{\rm Ad}$. Now, suppose we are given a cylindrical function
$f'$ on $\AGbar$. Its projection under the appropriate $\pi(\Sstar)$
is, by the preceeding result, contained in the projection of some
element of $\HAbar$. Thus, $\c$ is contained in $\HAbar$.

Combining the two results, we have the desired result: the $C^\star$
algebras $\HAbar$ and $\cbar$ are isomorphic.
\end{proof}

\begin{followon}{Remark}
Theorem 4.3 suggests that from the very beginning we could have
introduced the \C* $\cbar$ in place of $\HAbar$. This is indeed the case.
More precisely, we could have proceeded as follows. Begin with the
space $\A/\G$, introduce the notion of hoops, and hoop decomposition
in terms of independent hoops. Then given a subgroup $\Sstar$
generated by $n$ independent hoops, we could have introduced an
equivalence relation on $\AG$ ({\it not} $\AGbar$ which we do not yet
have!) as follows: $A_1 \sim A_2$ iff $H({\gt}, A_1) = g^{-1}H({\gt},
A_2)g$ for all $\gt \in \Sstar$ and some (hoop independent) $g\in SU(2)$.
It is then again true (due to Lemma 3.3) that the quotient $[\AG]
/\!\sim$ is isomorphic to $[SU(2)]^n/{\rm Ad}$. (This is true inspite
of the fact that we are using $\AG$ rather than $\AGbar$ because, as
remarked immediately after Lemma 4.1, each $\{\bar{A}\} \in
[\AGbar]/\!\sim$ contains a regular connection.) We can therefore
define cylindrical functions, but now on $\AG$ (rather than $\AGbar$)
as the pull--backs of continuous functions on $[SU(2)]^n/{\rm Ad}$.
These functions have a natural \C* structure.  We can use it as the
starting point in place of the A--I holonomy algebra.  While this
strategy seems natural at first from an analysis viewpoint, it has two
drawbacks, both stemming from the fact that the Wilson loops are now
assigned a secondary role. For physical applications, this is
unsatisfactory since Wilson loops are the basic observables of gauge
theories. From a mathematical viewpoint, the relation between
knot/link invariants and measures on the spectrum of the algebra would
now be obscure.  Nonetheless, it is good to keep in mind that this
alternate strategy {\it is} available as it may make other issues more
transparent.
\end{followon}

\subsection{A natural measure}
In this sub--section, we will discuss the issue of integration of
cylindrical functions on $\AGbar$. Our main objective is to introduce
a natural, faithful, diffeomorphism invariant, cylindrical measure on
$\AGbar$.

The idea is similar to the one used in the case of topological vector
spaces discussed in the beginning of the previous sub--section. Thus,
for each $n$, we wish to equip the spaces $[SU(2)]^n/{\rm Ad}$ with a
measure $d\tilde\mu(n)$ and, as in (4.1), to set
$$\int_{\AGbar}\ d\mu\ f := \int_{[SU(2)]^n/{\rm ad}}\
\ d\tilde\mu{(n)} \ \tilde{f}. \eqno(4.3)$$
It is clear that for the integral to be well--defined, the set of
measures we choose on $[SU(2)]^n/{\rm Ad}$ should be compatible so
that the analog of (4.2) holds. We will now exhibit a natural choice
for which the compatibility is automatically satisfied. Denote by
$d\mu$ the normalized Haar measure on $SU(2)$. It naturally induces a
measure on $[SU(2)]^n$ which is invariant under the adjoint action of
$SU(2)$ and therefore projects down unambiguously to $[SU(2)]^n/{\rm
Ad}$. This is our choice of $d\tilde\mu{(n)}$. This is a natural
choice since $SU(2)$ comes equipped with the Haar measure. In
particular, we have not introduced any additional structure on the
underlying 3--manifold $\S$ (on which the initial connections $A$ are
defined); hence the resulting cylindrical measure will be
automatically invariant under the action of $Diff(\S)$.

We will first show that the measures $d\tilde\mu{(n)}$ satisfy the
appropriate compatibility conditions and then explore the properties
of the resulting cylindrical measure on $\AGbar$. For simplicity of
presentation, in what follows we shall not draw a distinction between
functions on $[SU(2)]^n/{\rm Ad}$ and those on $[SU(2)]^n$ which are
invariant under the adjoint $SU(2)$--action. Thus, instead of carrying
our integrals over $[SU(2)]^n/{\rm Ad}$ of functions thereon, we will
often integrate their lifts to $[SU(2)]^n$.

\begin{theorem}
a) If $f$ is a cylindrical function on $\AGbar$ with respect to two
different finitely generated subgroups $\Sstar_i,\ i=1,2$ of the hoop
group, and $\tilde{f}_i$, its projections to $[SU(2)]^{n_i}/{\rm Ad}$,
then we have:
$$\int_{[SU(2)]^{n_1}/{\rm Ad}}\ d\tilde\mu{(n_1)}\ \tilde{f}_1 =
\int_{[SU(2)]^{n_2}/{\rm Ad}}\ d\tilde\mu{(n_2)}\ \tilde{f}_2; \eqno(4.4)$$
b) The functional $v: \HAbar\ \mapsto\ C$ on the A--I \C* $\HAbar$ defined
below is linear, strictly positive and Diff($\Sigma$)-invariant:
$$ v(F) = \ \ \int_{\AGbar}\ d\mu \ f \ ,\eqno(4.5)$$
where the cylindrical function $f$ on $\AGbar$ is the Gel'fand transform
the element $F$ of $\HAbar$; and

\noindent
c) the cylindrical `measure' $d\mu$ defined by (4.3) is a genuine, regular
and strictly positive measure on $\AGbar$; $d\mu$ is Diff($\S$)
invariant.
\end{theorem}

\begin{proof}: As in the proof of Theorem 4.3, let us first consider the $n_1$
generators of $\Sstar_1$ and $n_2$ generators of $\Sstar_2$ and use
the construction of Section 3.1 to carry out the decomposition of
these $n_1+n_2$ hoops in terms of independent hoops, say $\bt_1,
...,\bt_n$.  Thus, the original $n_1+n_2$ hoops are contained in the
group $\Sstar$ generated by the $\bt_i$. (Note that $n_i<n$ in the
non--trivial case when the original subgroups $\Sstar_1$ and
$\Sstar_2$ are distinct.) Clearly, $\Sstar_i\subset \Sstar$. Hence,
the given $f$ is cylindrical also with respect to $\Sstar$, i.e., is
the pull--back of a function $\tilde{f}$ on $[SU(2)]^n/{\rm Ad}$. We
will now establish (4.4) by showing that each of the two integrals in
that equation equals the analogous integral of $\tilde{f}$ over
$[SU(2)]^n/{\rm Ad}$.

To deal with quantities with suffixes 1 and 2 simultaneously, for
notational simplicity we will let a prime stand for {\it either} the
suffix 1 {\it or} 2.  Then, in particular, we have finitely generated
subgroups $\Sstar$ and $(\Sstar)'$ of the hoop group, with
$(\Sstar)'\subset \Sstar$, and a function $f$ which is cylindrical
with respect to both of them. The generators of $\Sstar$ are $\bt_1,
..., \bt_n$ while those of $(\Sstar)'$ are ${\bt_1}', ...,
{\bt_{n'}}'$ (with $n'<n$). From the construction of $\Sstar$ (Section
3.1) it follows that every hoop in the second set can be expressed as
a composition of the hoops in the first set and their inverses.
Furthermore, since the hoops in each set are strongly independent, the
construction implies that for each $i'\in \{1,..., n'\}$, there exists
$K(i') \in \{1,..., n\}$ with the following properties: i) if $i'\not=
j'$, $K(i')\not= K(j')$; and, ii) in the decomposition of
${\bt_{i'}}'$ in terms of unprimed hoops, the hoop $\bt_{K(i')}$ (or
its inverse) appears exactly once and if $i'\not= j'$, the hoop
$\bt_{K(j')}$ does not appear at all. For simplicity, let us assume
that the orientation of the primed hoops is such that it is
$\b_{K(i')}$ rather than its inverse that features in the
decomposition of ${\b_{i'}}'$.  Next, let us denote by $(g_1,...,g_n)$
a point of the quotient $[SU(2)]^n$ which is projected to $(g'_1,
...,g'_{n'})$ in $[SU(2)]^{n'}$. Then $g'_{i'}$ is expressed in terms
of $g_i$ as $g'_{i'} = ..g_{K(i')}..$, where $..$ denotes a product of
elements of the type $g_m$ where $m \not= K(j)$ for any $j$. Since the
given function $f$ on $\AGbar$ is a pull--back of a function
$\tilde{f}$ on $[SU(2)]^n$ as well as of a function ${\tilde{f}}'$ on
$[SU(2)]^{n'}$, it follows that:

\begin{displaymath}
{\tilde{f}} (g_1, ..., g_n)
={\tilde{f}}' (g'_1, ..., g'_{n'})
   = {\tilde{f}}' (..g_{K(1)}.., ..., ..g_{K(n')}..).
   \end{displaymath}
Hence,
\begin{eqnarray}
&{}&\int d\tilde{\mu}_1 ... \int d\tilde{\mu}_n \
{\tilde{f}} (g_1, ..., g_n) = \nonumber \\
&{}&\int\prod_{m\not=K(1)...K(n')}d\tilde{\mu}_m \big( \int
d\tilde{\mu}_{K(1)} ... \int d\tilde{\mu}_{K(n')}
{\tilde{f}}' (..g_{K(1)}.., ..., ..g_{K(n')}..)\ \big) \nonumber\\
&{}&= \int \prod_{m=n'+1}^{n}d\tilde{\mu}_m \big(\int d\tilde{\mu}_{1}
... \int d\tilde{\mu}_{n'} {\tilde{f}}' (g_{1}, ..., g_{n'})\ \big)
\nonumber \\
&{}&= \int d\tilde{\mu}_{1} ... \int d\tilde{\mu}_{n'}
{\tilde{f}}' (g_{1}, ..., g_{n'}),\nonumber
\end{eqnarray}
where, in the first step, we simply rearranged the order of integration, in
the second step we used the invariance property of the Haar measure and
simplified notation for dummy variables being integrated, and,
in the third step, we have used the fact that, since the measure is
normalized, the integral of a constant produces just that constant%
\footnote{In the simplest case, with $n=2$ and $n'=1$, we would
for example have $\tilde{f}(g_1,g_2) = \tilde{f}'(g_1\cdot g_2) \equiv
\tilde{f}'(g')$. Then $\int d\mu(g_1)\int d\mu(g_2) \tilde{f}(g_1, g_2) =
\int d\mu(g_1)\int d\mu(g_2) \tilde{f}'(g_1\cdot g_2) = \int d\mu(g')
\tilde{f}(g')$, where in the second step we have used the invariance and
normalization properties of the Haar measure.}.
This establishes the required result (4.4).

Next, consider the ``vaccum expectation value functional'' $v$ on
$\HAbar$ defined by (4.5).
%defines on ${\cal C}$ ($\equiv \HAbar$) a linear functional
%$v: \HAbar \mapsto {\bf C}$  via the ``vacuum
%expectation values'': $v(f) = \int d\mu \ f$.
The continuity and linearity of $v$ follow directly from its definition.
That it is positive, i.e., satisfies $v({F}^\star F) \ge 0$ is equally
obvious. In fact, a stronger result holds: if $F \not= 0$, then $v(F^\star
F)>0$ since the value $v(F^\star F)$ of $v$ on $F^\star F$ is obtained
by integrating a non--negative function $(\tilde f)^\star \tilde{f}$
on $[SU(2)]^n/{\rm Ad}$ with respect to a regular measure. Thus,
$v$ is {\it strictly} positive.

To see that we have a regular measure on $\AGbar$, recall
first that the \C* of cylindrical functions is naturally isomorphic
with the Gel'fand transform of the \C* $\HAbar$, which, in turn is the
\C* of all continuous functions on the compact, Hausdorff space
$\AGbar$. The ``vacuum expectation value function'' $v$ is therefore a
continuous linear functional on the algebra of all continuous
functions on $\AGbar$ equipped with the $L^\infty$ (i.e., sup--)norm.
Hence, by the Reisz--Markov theorem, it follows that $v(f) \equiv \int
d\mu f$ is in fact the integral of the continuous function $f$ on
$\AGbar$ with respect to a regular measure.

Finally, since our construction did not require the introduction of
any extra structure on $\S$ (such as a metric or a volume element), it
follows that the functional $v$ and the measure $d\mu$ are
Diff($\S$) invariant.
\end{proof}

We will refer to $d\mu$ as the ``induced Haar measure'' on $\AGbar$.
We now explore some properties of this measure and of the resulting
representation of the holonomy \C* $\HAbar$.

First, the representation of $\HAbar$ can be constructed as follows:
The Hilbert space is $L^2 (\AGbar ,d\mu)$ and the elements $F$ of
$\HAbar$ act by multiplication so that we have $F \Psi = f\cdot\Psi$
for all $\Psi\in L^2(\AGbar, d\mu)$, where $f \in {\cal C}$
is the Gel'fand transform of $F$. This representation is cyclic and
the function $\Psi_o$ defined by $\Psi_o (\Ab) = 1$ on $\AGbar$ is the
``vacuum'' state.  Every generator $T_\a$ of the holonomy \C* is
represented by a bounded, self--adjoint operator on $L^2(\AGbar,
d\mu)$.

Second, it follows from the strictness of the positivity of the
measure that the resulting representation of the \C* $\HAbar$ is {\it
faithful}. To our knowledge, this is the first faithful representation
of the holonomy \C* that has been constructed explicitly. Finally, the
diffeomorphism group Diff($\S$) of $\Sigma$ has an induced action on
the \C* $\HAbar$ and hence also on its spectrum $\AGbar$. Since the
measure $d\mu$ on $\AGbar$ is invariant under this action,
the Hilbert space $L^2(\AGbar, d\mu)$ carries a unitary representation
of Diff($\S$). Under this action, the ``vacuum state'' $\Psi_0$ is
left invariant.

Third, restricting the values of $v$ to the generators $T_{\at}\equiv
[\a]_K$ of $\HAbar$, we obtain the generating functional $\Gamma(\a)$
of this representation: $\Gamma (\a):= \int d\mu \bar{A}(\at)$. This
functional on the loop space $\L$ is diffeomorphism invariant since
the measure $d\mu$ is.  It thus defined a generalized knot invariant.
We will see below that, unlike the standard invariants, $\Gamma(\a)$
vanishes on all smoothly embedded loops. It's action is non--trivial
only when the loop is re--traced, i.e. only on generalized loops.

 Let us conclude this section by indicating how one computes the
integral in practice. Fix a loop $\a \in \L$ and let
  $$[\a] =[\b_{i_1}]^{\epsilon_1}...[\b_{i_N}]^{\epsilon_N}\eqno(4.6)$$
be the minimal decomposition of $\a$ into (strongly) independent loops
$\b_1$, ...,$\b_n$ as in Section 3.1, where $\epsilon_k\in \{-1,1\}$
keeps track of the orientation and $i_k\in \{1, ..., n\}$. (We have
used the double suffix $i_k$ because any one independent loop,
$\b_1,...,\b_n$ can arise more than once in the decomposition. Thus,
$N\ge n$.) The loop $\a$ defines an element $T_{\at}$ of the holonomy
\C* $\HAbar$ whose Gel'fand transform $t_{\at}$ is a function on
$\AGbar$. This is the cylindrical function we wish to integrate. Now,
the projection map of Lemma 4.1 assigns to $t_{\at}$ the function
$\tilde{t}_{\at}$ on $[SU(2)]^n/{\rm Ad}$ given by:
$$\tilde {t}_{\at}(g_1,...,g_n) := \Tr\ (g_{i_1}^{\epsilon_1}...
g_{i_N}^{\epsilon_N}).$$
Thus, to integrate $t_{\at}$ on $\AGbar$, we need to integrate this
$\tilde{t}_{\at}$ on $[SU(2)]^n/{\rm Ad}$. Now, there exist in the
literature useful formulas for integrating polynomials of the ``matrix
element functions'' over $SU(2)$ (see, e.g., [14] ). These can be now
used to evaluate the required integrals on $[SU(2)]^n/{\rm Ad}$. In
particular, one can readily show the following result: $$\int_{\AGbar}
d\mu\ t_{\at} = 0, $$ unless {\it every} loop $\b_i$ is repeated an
even number of times in the decomposition (4.6) of $\a$.

\section{Discussion}

In this paper, we completed the A--I program in several directions.
First, by exploiting the fact that the loops are piecewise analytic,
we were able to obtain a complete characterization of the Gel'fand
spectrum $\AGbar$ of the holonomy \C* $\HAbar$. A--I had shown that
every element of $\AGbar$ defines a homomorphism from the hoop group
$\HG$ to $SU(2)$. We found that every homomorphism from $\HG$ to
$SU(2)$ defines an element of the spectrum and two homomorphisms
$\hat{H}_1$ and $\hat{H}_2$ define the same element of $\AGbar$ only
if they differ by an $SU(2)$ automorphism, i.e., if and only if
$\hat{H}_2 = g^{-1}\cdot\hat{H}_1\cdot g$ for some $g\in SU(2)$. Since
this characterization is rather simple and purely algebraic, it is
useful in practice. The second main result also intertwines the
structure of the hoop group with that of the Gel'fand spectrum. Given
a subgroup $\Sstar$ of $\HG$ generated by a finite number, say $n$, of
(strongly) independent hoops, we were able to define a canonical
projection $\pi(\Sstar)$ from $\AGbar$ to the compact space
$[SU(2)]^n/{\rm Ad}$. This family of projections enables us to define
cylindrical functions on $\AGbar$: these are the pull--backs to
$\AGbar$ of the continuous functions on $[SU(2)]^n/{\rm Ad}$.  We
analysed the space ${\cal C}$ of these functions and found that it has
the structure of a \C* which, moreover, is naturally isomorphic with
the \C* of all continuous functions on the compact Hausdorff space
$\AGbar$ and hence, via the inverse Gel'fand transform, also to the
holonomy \C* $\HAbar$ with which we began. We then carried out a
non--linear extension of the theory of cylindrical measures to
integrate these cylindrical functions on $\AGbar$.  Since ${\cal C}$
{\it is} the \C* of all continuous functions on $\AGbar$, the
cylindrical measures are in fact regular measures on $\AGbar$.
Finally, we were able to introduce explicitly such a measure on
$\AGbar$ which is natural in the sense that it arises from the Haar
measure on $SU(2)$.  The resulting representation of the holonomy \C*
has several interesting properties. In particular, the representation
is faithful and diffeomorphism invariant. Hence, its generating
function $\Gamma$---which sends loops $\a$ (based at $\xz$) to real
numbers $\Gamma(\a)$---defines a generalized knot invariant
(generalized because we have allowed loops to have kinks,
self--intersections and self--overlaps). These constructions show that
the A--I program can be realized in detail.

Having the induced Haar measure $d\mu$ on $\AGbar$ at one's disposal,
we can now make the ideas on loop transform $\T$ of [3,4] rigorous. The
transform $\T$ sends states in the connection representation to states
in the so--called loop representation. In terms of machinery
introduced in this paper, a state $\Psi$ in the connection
representation is a square--integrable function on $\AGbar$; thus
$\Psi \in L^2(\AGbar, d\mu)$. The transform $\T$ sends it to a
function $\psi$ on the space $\HG$ of hoops.  (Sometimes, it is
convenient to lift $\psi$ canonically to the space $\L$ and regard it
as a function on the loop space. This is the origin of the term ``loop
representation.'') We have: $\T\circ\Psi = \psi$ with
$$ \psi(\at) = \int_{\AGbar} d\mu \ \overline{t_{\at}(\Ab)}
\Psi (\Ab)  =<t_{\at}, \Psi>, \eqno(5.1)$$
where, $t_{\at}$ (with $t_{\at}(\Ab) = \Ab(\at)$) is the Gel'fand
transform of the trace of the holonomy function, $T_{\at}$, on $\AG$,
associated with the hoop $\at$, and $<.,.>$ denotes the inner
product on $L^2(\AGbar, d\mu)$. Thus, in the transform the traces of
holonomies play the role of the ``integral kernel.'' Elements of the
holonomy algebra have a natural action on the connection states
$\Psi$. Using $\T$, one can transform this action to the hoop states
$\psi$. It turns out that the action is surprisingly simple and can be
represented directly in terms of elementary operations on the hoop
space. Consider a generator $T_{\gt}$ of $\HA$.  In the connection
representation, we have: $(T_{\gt}\circ \Psi)(\Ab) =
T_{\gt}(\Ab)\cdot\psi(\Ab)$. On the hoop states, this action
translates to:
$$(T_{\gt}\circ \psi)(\at) = {1\over 2}(\psi(\at\cdot
\gt) + \psi(\at \cdot {\gt}^{-1})). \eqno(5.2)$$
where $\at\cdot\gt$ is the composition of hoops $\at$ and $\gt$ in the
hoop group. Thus, one can forgo the connection representation and work
directly in terms of hoop states. We will show elsewhere that the
transform also interacts well with the ``momentum operators'' which
are associated with closed strips, i.e. that these operators also have
simple action on the hoop states.

The hoop states are especially well--suited to deal with
diffeomorphism (i.e.\ Diff($\S$)) invariant theories. In such
theories, physical states are required to be invariant under
Diff($\S$). This condition is awkward to impose in the connection
representation and, it is difficult to control the structure of the
space of resulting states. In the loop representation, by contrast,
the task is rather simple: Physical states depend only on generalized
knot and link classes of loops. Because of this simplification and
because the action of the basic operators can be represented by simple
operations on hoops, the loop representation has proved to be a
powerful tool in quantum general relativity in 3 and 4 dimensions.

The use of this representation, however, raises several issues which
are still open. Perhaps the most basic of these is that, without
referring back to the connection representation, we do not yet have a
useful characterization of the hoop states which are images of
connection states, i.e. of elements of $L^2(\AGbar, d\mu)$. Neither do
we have an expression of the inner product between hoop states. It
would be extremely interesting to develop integration theory also over
the hoop group $\HG$ and express the inner product between hoop states
directly in terms of such integrals, without any reference to the
connection representation. This may indeed be possible using again the
idea of cylindrical functions, but now on $\HG$, rather than on
$\AGbar$ and exploiting, as before, the duality between these spaces.
If this is achieved and if the integrals over $\AGbar$ and $\HG$ can
be related, we would have a non--linear generalization of the
Plancherel theorem which establishes that the Fourier transform is an
isomorphism between two spaces of $L^2$--functions. The loop transform
would then become an extremely powerful tool.

\section*{Appendix A:\ \ \ \ \ $C^1$ loops and $U(1)$--connections}
In the main body of the paper, we restricted ourselves to piecewise
analytic loops. This restriction was essential in Section 3.1 for our
decomposition of a given set of finite number of loops into {\it
independent} loops which in turn is used in every major result
contained in this paper. The restriction is not as severe as it might
first seem since every smooth 3--manifold admits an unique analytic
structure. Nonetheless, it is important to find out if our arguments
can be replaced by more sophisticated ones so that the main results
would continue to hold even if the holonomy \C* were constructed from
piecewise smooth loops. In this appendix, we consider $U(1)$
connections (rather than $SU(2)$) and show that, in this theory, our
main results do continue to hold for piecewise $C^1$ loops. However,
the new arguments make a crucial use of the Abelian character of
$U(1)$ and do not by themselves go over to non--Abelian theories.
Nonetheless, this Abelian example is an indication that analyticity
may not be indispensible even in the non--Abelian case.

The appendix is divided into three parts. The first is devoted to
certain topological considerations which arise because 3--manifolds
admit non--trivial $U(1)$--bundles. The second proves the analog of
the spectrum theorem of Section 3. The third introduces a
diffeomorphism invariant measure on the spectrum and discusses some of
its properties. Throughout this appendix, by loops we shall mean
continuous, piecewise $C^1$ loops. We will use the same notation as in
the main text but now those symbols will refer to the $U(1)$ theory.
Thus, for example, $\HG$ will denote the $U(1)$--hoop group and $\HA$,
the $U(1)$ holonomy $\star$--algebra.

\subsection*{A.1\ \ \ \ \  Topological considerations}
Fix a smooth 3--manifold $\S$. Denote by $\A$ the space of all smooth
$U(1)$ connections which belong to an appropriate Sobolev space. Now,
unlike $SU(2)$ bundles, $U(1)$ bundles over 3--manifolds need not be
trivial.  The question therefore arises as to whether we should allow
arbitrary $U(1)$ connections or restrict ourselves only to the trivial
bundle in the construction of the holonomy \C*. Fortunately, it turns
out that both choices lead to the same \C*. This is the main result of
this sub--section.

Denote by $\A^o$, the sub--space of $\A$ consisting of connections on
the trivial $U(1)$ bundle over $\S$. As in common in the physics
literature, we will identify elements $A^o$ of $\A^o$ with
real--valued 1--forms. Thus, the holonomies defined by any $A^0\in
\A^0$ will be denoted by: $H(\a, A^o)= \exp (i\oint_\a A^o) \equiv
\exp (i\theta)$, where $\theta$ takes values in $(0, 2\pi)$ and
depends on both the connection $A^o$ and the loop $\a$. We begin with
the following result:

\medskip
\noindent
{\bf Lemma A.1} \ {\em Given a finite number of loops, $\a_1,
...,\a_n$, for every $A\in\A $, there exists $A^o\in {\A}^o $ such
that:}
$$H(\a_i, A) = H(\a_i, A^o), \quad \forall i \in\{1,...,n\}.$$
\medskip

\begin{proof}
Suppose the connection $A$ is defined on the bundle $P$.  We will show
first that there exists a local section $s$ of $P$ which contains the
given loops $\a_i$. To construct the section we use a map
$\Phi:\S\rightarrow S_2$ which generates the bundle $P(\S,U(1))$ (see
Trautman [15]). The map $\Phi$ carries each $\a_i$ into a loop
${\check \a_i}$ in $S_2$. Since the loops ${\check \a_i}$ can not form
a dense set in $S_2$, we may remove from the 2--sphere an open ball
$B$ which does not intersect any of ${\check \a_i}$.  Now, since
$S_2-B$ is contractible, there exists a smooth section $\sigma$ of the
Hopf bundle $U(1)\rightarrow SU(2) \rightarrow S_2$, $$\sigma:
(S_2-B)\mapsto SU(2).$$ Hence, there exists a section $s$ of $P(\S,
U(1))$ defined on a subset $V=\Phi^{-1}(S_2-B)\subset\S$ which
contains all the loops $\a_i$. Having obtained this section $s$, we
can now look for the connection $A^o$. Let $\omega$ be a globally
defined, real 1--form on $\S$ such that $\omega|V = s^*A.$ Hence, the
connection $A^0:= \omega$ defines on the given loops $\a_i$ the same
holonomy elements as $A$; i.e.  $$H(\a_i, A^0)=H(\a_i,A)\eqno(A.1)$$
for all $i$.
\end{proof}

This lemma has several useful implications. We mention two that will
be used immediately in what follows.

\begin{enumerate}
\item Definition of hoops: A priori, there are two equivalence relations
on the space $\L$ of based loops that one can introduce. One may say
that two loops are equivalent if the holonomies of all connections in
$\A$ around them are the same, or, one could ask that the holonomies
be the same only for connections in $\A^o$. Lemma A.1 implies that the
two notions are in fact equivalent; there is only one hoop group
$\HG$. As in the main text, we will denote by $\at$ the hoop to which
a loop $\a\in\L$ belongs.
\item Sup norm: Consider functions $f$ on $\A$ defined by finite linear
combinations of holono\-mi\-es on $\A$ around hoops in $\HG$. (Since the
holonomies themselves are complex numbers, the trace operation is now
redundant.) We can restrict these functions to $\A^o $. Lemma A.1 implies that:
$$\sup_{A\in\A } |f(A)| = \sup_{A^o\in\A^o} |f(A^o)|\eqno(A.2)$$
\end{enumerate}

As a consequence of these implications, we can use {\it either} $\A$
{\it or} $\A^o$ to construct the holonomy \C*; inspite of topological
non--trivialities, there is only one $\HAbar$. To construct this
algebra we proceed as follows. Let $\HA$ denote the complex vector
space of functions $f$ on $\A^o$ of the form
$$f(A^o) = \sum_{j=1}^n a_j H(\a, A^o) \equiv \sum_{j=1}^n a_j
\exp (i\oint_\a A^o) \ ,\eqno(A.3)$$
where $a_j$ are complex numbers. Clearly, $\HA$ has the
structure of a $\star$--algebra with the product law $H(\a , A^o)\circ
H(\b , A^o) = H(\a\cdot \b , A^o)$ and the $\star$--relation given by
$(H(\a, A^o))^\star = H(\a^{-1}, A^o)$. Equip it with the sup--norm
and take completion. The result is the required \C* $\HAbar$.

We conclude by noting another consequence of Lemma A.1: the (Abelian)
hoop group $\HG$ has no torsion. More precisely, we have the following
result:

\medskip \noindent
{\bf Lemma A.2} \ {\em
Let $\at \in \HG $. Then if $(\at )^n = e$, the identity in $\HG$, for
$n\in {\bf Z}$, then $\at = e$.}
\medskip

\begin{proof} Since $(\at )^n =e$, we have, for every $A^o\in \A^o$,
$$ H (\a , A^o) \equiv H(\a^n, \textstyle{1\over n} A^o) = 1, $$
where $\a$ is any loop in the hoop $\at$. Hence, by Lemma A.1, it follows
that $\at = e$.
\end{proof}

Although the proof is more complicated, the analogous result holds also
in the $SU(2)$-case treated in the main text. However, in that case, the
hoop group is non-Abelian. As we will see below, it is the Abelian character
of the $U(1)$-hoop group that makes the result of this lemma useful.

\subsection*{A.2\ \ \ \ \  The Gel'fand spectrum of $\HAbar$}
We now want to obtain a complete characterization of the spectrum of
the $U(1)$ holonomy \C* $\HAbar$ along the lines of Section 3. The
analog of Lemma 3.2 is easy to establish: Every element $\Ab$ of the
spectrum defines a homomorphism $\hat{H}_{\Ab}$ from the hoop group
$\HG$ to $U(1)$ (now given simply by $\hat{H}_{\Ab} (\at) = \Ab
(\at)$). This is not surprising. Indeed, it is clear from the
discussion in Section 3 that this lemma continues to hold for
piecewise smooth loops even in the full $SU(2)$ theory. However, the
situation is quite different for Lemma 3.3 because there we made a
crucial use of the fact that the loops were (continuous and) piecewise
{\it analytic}. Therefore, we must now modify that argument suitably.
An appropriate replacement is contained in the following lemma.

\medskip\noindent
{\bf Lemma A.3} \ {\em For every homomorphism $\hat{H}$ from the hoop
group $\HG$ to $U(1)$, every finite set of hoops $\{\at_1,
...,\at_k\}$ and every $\epsilon>0$, there exists a connection $A^o\in
\A^o $ such that:}
   $$ |\hat{H}(\at_i) - H(\a_i , A^o)| < \epsilon ,\ \ \forall i\in
       \{1,...,k\}\eqno(A.4)$$
\medskip

\begin{proof}
Consider the subgroup $\HG(\at_1,...,\at_n)\subset\HG$ generated by
$\at_1$,... ...,$\at_k$. Since $\HG$ is Abelian and since it has no
torsion, $\HG(\at_1, ...,\at_k)$ is finitely and freely generated by
some elements, say $\bt_1, ...,\bt_n$. Hence, if $(A.4)$ is satisfied
by ${\bt_i}$ for a sufficiently small $\epsilon'$ then it will also be
satisfied by the given $\at_j$ for the given $\epsilon$. Consequently,
without loss of generality, we can assume that the hoops $\a_i$ are
(weakly) $U(1)$--independent, i.e., that they satisfy the following
condition:
$${\rm if}\ \ (\at_1)^{k_1}...(\at_n)^{k_n} = e, \quad {\rm
then},\ \ k_i = 0\ \forall i.$$

Now, given such hoops $\at_i$, the homomorphism $\hat{H}: \HG
\mapsto U(1)$ defines a point $(\hat{H}(\at_1), ..., \hat{H}(\at_k))
\in [U(1)]^k$. On the other hand, there also exists a map from $\A^o$
to $[U(1)]^k$, which can be expressed as a composition:
$$\A^o \ni A^o\mapsto  (\oint_{\a_1}A^o,..., \oint_{\a_k}A^o)
\mapsto (e^{i\oint_{\a_1}A^o},..., e^{i\oint_{\a_k}A^o})\in
[U(1)]^k. \eqno(A.5)$$
The first map in $(A.5)$ is linear and its image, $V\subset R^k$, is a
vector space. Let $0\not={\bf m}=(m_1,...,m_k)\in {\bf Z}^k$. Denote
by $V_{\bf m}$ the subspace of $V$ orthogonal to ${\bf m}$. Now if
$V_{\bf m}$ were to equal $V$ then we would have
  $$(\at_1)^{m_1}...(\at_k)^{m_k}=1, $$
which would contradict the assumption that the given set of hoops is
independent. Hence we necessarily have: $V_{\bf m}<V$. Furthermore,
since a countable union of subsets of measure zero has measure zero,
it follows that
$$\bigcup_{{\bf m}\in {\bf Z}^k} V_{\bf m}\ <\ V .$$ This strict
inequality implies that there exists a connection $B^o\in \A^o$ such
that
 $${\bf v} := (\oint_{\a_1}B^0,..., \oint_{\a_n}B^0)\
\in\ (V - \bigcup_{{\bf m}\in {\bf Z}^k} V_{\bf m}).$$
In other words, $B^o$ is such that every ratio $v_i/v_j$ of two
different components of ${\bf v}$ is irrational. Thus, the line in
$\A^o$ defined by $B^0$ via $A^o(t) = tB^o$ is carried by the map
$(A.5)$ into a line which is dense in $[U(1)]^k$. This ensures that
there exists an $A^o$ on this line which has the required property
$(A.4)$.
\end{proof}

Armed with this substitute of Lemma 3.3, it is straightforward to show
the analog of Lemma 3.4. Combining these results, we have the $U(1)$
spectrum theorem:

\medskip
{\bf Theorem A.4} \ {\em
Every $\Ab$ in the spectrum of $\HAbar$ defines a homomorphism
$\hat{H}$ from the hoop group $\HG$ to $U(1)$, and, conversely, every
homomorphism $\hat{H}$ defines an element $\Ab$ of the spectrum such
that $\hat{H}(\at)= \Ab(\at)$. This is a 1--1 correspondence.}
\medskip

\subsection*{A.3\ \ \ \ \  A natural measure}
Results presented in the previous sub--section imply that one can
again introduce the notion of cylindrical functions and measures. In
this sub--section, we will exhibit a natural measure. Rather than
going through the same constructions as in the main text, for the sake
of diversity, we will adopt here a complementary approach: We will
present a (strictly--) positive linear functional on the holonomy \C*
$\HAbar$ whose properties suffice to ensure the existence of a
diffeomorphism invariant, faithful, regular measure on the spectrum of
$\HAbar$.

We know from the general theory outlined in Section 2 that, to specify
a positive linear functional on $\HAbar$, it suffices to provide an
appropriate generating functional $\Gamma [\a]$ on $\L$. Let us simply
set
$$\Gamma (\a) = \cases {1,& if $\at = \tilde{o}$ \cr
0, & otherwise\cr }, \eqno(A.6)$$
where $\tilde{o}$ is the identity hoop in $\HG$.  It is clear that
this functional on $\L$ is diffeomorphism invariant. Indeed, it is the
simplest of such functionals on $\L$. We will now show that it does
have all the properties to be a generating function
\footnote{Since this functional is so simple and natural, one might
imagine using it to define a positive linear functional also for the $SU(2)$
holonomy algebra of the main text. However, this strategy fails
because of the $SU(2)$ identities. For example, in the $SU(2)$ holonomy
algebra we have $T_{\a\cdot\b\cdot\b\cdot\a} + T_{\a\cdot\b\cdot\a^{-1}
\cdot\b^{-1}} -T_{\a\cdot\b\cdot\a\cdot\b} = 1$. Evaluation of the generating
functional $(A.6)$ on this identity would lead to the contradiction $0=1$. We
{\it can} define a positive linear function on the $SU(2)$ holonomy \C* via:
$$\Gamma'(\a) =\cases{1, & if $T_{\at}(A) = 1 \forall A\in \A^o$\cr
                        0, & otherwise,\cr}$$
where, we have regarded $\A^o$ as a subspace of the space of $SU(2)$
connections. However, on the $SU(2)$ holonomy algebra, this positive linear
functional is not strictly positive: the measure on $\AGbar$ it defines is
concentrated just on $\A^o$. Consequently,  the resulting representation of
the $SU(2)$ holonomy \C* fails to be faithful.}.

{\bf Theorem A.5} \ {\em
$\Gamma(\a)$ extends to a continuous positive linear
function on the holonomy \C* $\HA$. Furthermore, this function is
strictly positive.}

\begin{proof}
By its definition, the loop functional $\Gamma (\a)$ admits a
well--defined projection on the hoop space $\HG$ which we will denote
again by $\Gamma$. Let ${\cal FHG}$ be the free vector space generated
by the hoop group. Extend $\Gamma$ to ${\cal FHG}$ by linearity.  We
will first establish that this extension satisfies the following
property: Given a finite set of hoops $\at_j, j =1,...,n$ such that,
if $\at_i \not=\at_j$ if $i\not= j$, we have
$$ \Gamma \big((\sum_{j=1}^{n} a_j \at_j)^\star (\sum_{j=1}^{n} a_j
\at_j)\big)  < \sup_{A^o\in\A^o} (\sum_{j=1}^{n} a_j H(\at_j, A^o))^\star
(\sum_{j=1}^{n} a_j H(\at_j, A^o)). \eqno(A.7)$$
To see this, note first that, by definition of $\Gamma$, the left side
equals $\sum |a_j|^2$. The right side equals $(\sum a_j \exp
i\theta_j)^\star (\sum a_j \exp i\theta_j)$ where $\theta_j$ depend on
$A^o$ and the hoop $\at_j$. Now, by Lemma A.3, given the $n$ hoops
$\at_j$, one can find a $A^o$ such that the angles $\theta_j$ can be
made as close as one wishes to a pre--specified n--tuple $\theta^o_j$.
Finally, given the $n$ complex numbers $a_i$, one can always find
$\theta^o_j$ such that $\sum |a_j|^2 < |(\sum a_j \exp i\theta^o_j)^\star
(\sum a_j \exp i\theta^o_j)|$.  Hence we have $(A.7)$.

The inequality $(A.7)$ implies that the functional $\Gamma$ has a
well--defined projection on the holonomy $\star$--algebra $\HA$ (which
is obtained by taking the quotient of ${\cal FHG}$ by the subspace $K$
consisting of $\sum b_i
\at_i$ such that $\sum b_i H(\at_i, A^o) = 0$ for all $A^o\in \A^o$.)
Furthermore, the projection is positive definite; $\Gamma(f^\star f) \ge 0$
for all $f\in \HA$, equality holding {\it only if} $f=0$. Next, since the
norm on this $\star$--algebra is the sup--norm on $\A^o$, it follows from
$(A.7)$ that the functional $\Gamma$ is continuous on $\HA$. Hence it admits
a unique continuous extension to the \C* $\HAbar$. Finally, $(A.7)$ implies
that the functional continues to be strictly positive on $\HAbar$.
\end{proof}
Theorem A.5 implies that the generating functional $\Gamma$ provides a
continuous, faithful representation of $\HAbar$ and hence a regular,
strictly positive measure on the Gel'fand spectrum of this algebra. It
is not difficult to verify by direct calculations that this is
precisely the $U(1)$ analog of the induced Haar measure discussed in
Section 4.

\section*{Appendix B:\ \ \ \  $C^\w$ loops, $U(N)$ and $SU(N)$--connections}
In this appendix, we will consider another extension of the results
presented in the main paper. We will now work with analytic manifolds
and piecewise analytic loops as in the main text. However, we will let
the manifold have any dimension and let the gauge group $G$ be either
$U(N)$ or $SU(N)$. Consequently, there are two types of complications:
algebraic and topological. The first arise because, e.g., the products
of traces of holonomies can no longer be expressed as sums while the
second arise because connections in question may be defined on
non--trivial principal bundles.  Nonetheless, we will see that the
main results of the paper continue to hold in these cases as well.
Several of the results also hold when the gauge group is allowed to be
any compact Lie group. However, for brevity of presentation, we will
refrain from making digressions to the general case. Also, since most
of the arguments are rather similar to those given in the main text,
details will be omitted.

Let us begin by fixing a principal fibre bundle $P(\S, G)$, obtain the
main results and then show, at the end, that they are independent of
the initial choice of $P(\S, G)$.Definitions of the space $\L$ of
based loops, and holonomy maps $H(\a, A)$, are the same as in Section
2. To keep the notation simple, we will continue to use the same
symbols as in the main text to denote various spaces which, however,
now refer to connections on $P(\S, G)$. Thus, $\A$ will denote the
space of (suitably regular) connections on $P(\S, G)$, $\HG$ will
denote the $P(\S,G)$--hoop group, and $\HAbar$ will be the holonomy
\C* generated by (finite sums of finite products of) traces of
holonomies of connections in $\A$ around hoops in $\HG$.

\subsection*{B.1\ \ \ \ \  Holonomy \C*}

We will set $T_{\at}(A)= \textstyle{1\over N}\Tr H(\a, A)$, where the
trace is taken in the fundamental representation of $U(N)$ or $SU(N)$,
and regard $T_{\at}$ as a function on $\AG$, the space of gauge
equivalent connections. The holonomy algebra $\HA$ is, by definition,
generated by all finite linear combinations (with complex
coefficients) of finite products of the $T_{\at}$, with the
$\star$--relation given by $f^\star = \bar{f}$, where the ``bar''
stands for complex--conjugation.  Being traces of $U(N)$ and $SU(N)$
matrices, $T_{\at}$ are bounded functions on $\AG$. We can therefore
introduce on $\HA$ the sup--norm $$||f|| = \sup_{A\in \A} |f(A)|,$$
and take the completion of $\HA$ to obtain a \C*. We will denote it by
$\HAbar$. This is the holonomy \C*.

The key difference between the structure of this $\HA$ and the one we
constructed in Section 2 is the following. In Section 2, the
$\star$--algebra $\HA$ was obtained simply by imposing an equivalence
relation ($K$, see Eq.\ (2.7)) on the free vector space generated by
the loop space $\L$. This was possible because, due to $SU(2)$
Mandelstam identities, products of any two $T_{\at}$ could be
expressed as a linear combination of $T_{\at}$ (Eq.\ (2.6)). For
groups under consideration, the situation is more involved. Let us
summarize the situation in the slightly more general context of the
group $GL(N)$. In this case, the Mandelstam identities follow from the
contraction of $N+1$ matrices ---$H(\a_1,A),...,H(\a_{N+1},A)$ in our
case--- with the identity
$$\delta_{[i_1}^{j_1},...,\delta_{i_{N+1}]}^{j_{N+1}}\ \ =\ \ 0$$
where $\delta_{i}^{j}$ is the Kronecker delta and the bracket stands
for the antisymmetrization. They enable one to express products of
$N+1$ $T_{\at}$--functions as a linear combination of traces of
products of $1, 2, ..., N$ $T_{\at}$--functions. Hence, we have to
begin by considering the free algebra generated by the hoop group and
then impose on it all the suitable identities by extending the
definition of the kernel (2.7). Consequently, if the gauge group is
$GL(N)$, an element of the holonomy algebra $\HA$ is expressible as
an equivalence class $[a\a, b \a_1\cdot \a_2,..., c \g_1\cdot...\g_{N}]$,
where $a,b,c$ are complex numbers; products involving $N+1$ and higher
loops are redundant. (For sub-groups of $GL(N)$ --such as
$SU(N)$--- further reductions may be possible.)

Finally, let us note identities that will be useful in what follows. These
result from the fact that the determinant of $N\times N$ matrix $M$, can
be expressed in terms of traces of powers of that
matrix, namely
$${\rm det}\ M\ \ =\ \ F(\Tr M, \Tr M^2,..., \Tr M^N)$$
for a certain polynomial $F$. Hence, if the gauge--group $G$ is
$U(N)$, we have, for any hoop $\at$,
$$F(T_{\at},...,T_{{\at}^N}) F(T_{{\at}^{-1}},...,T_{{\at}^{-N}})\
 =\ 1.\eqno(B.1a)$$
In the $SU(N)$ case a stronger identity holds:
$$F(T_{\at},...,T_{{\at}^N})\ \ = \ \ 1.\eqno(B.1b)$$

\subsection*{B.2 \ \ \ \ \  Loop decomposition and the spectrum theorem}
The construction of Section 3.1 for decomposition of a finite number
of loops into (strongly) {\it independent loops} makes no direct
reference to the gauge group and therefore carries over as is. Also,
it is still true that the map
$$\A\ \ni A\ \mapsto\ (H(\b_1,A),...,H(\b_n,A))\in G^n$$
is onto if the loops $\b_i$ are independent. (The proof requires some
minor modifications which consist of using local sections and a
sufficiently large number of generators of the gauge group). A direct
consequence is that the analog of Lemma 3.3 continues to hold.

As for the Gel'fand spectrum, the A--I result that there is a natural
embedding of $\AG$ into the spectrum $\AGbar$ of goes through as before
and so does the general argument due to Rendall [5] that the image of
$\AG$ is dense in $\AGbar$. (This is true for any group and representation
for which the traces of holonomies separate the points of $\AG$.) Finally,
using the analog of Lemma 3.3, it is straightforward to establish the analog
of the main conclusion of Lemma 3.4.

The converse ---the analog of the Lemma 3.2--- on the other
hand requires further analysis based on Giles' results along the lines
used by A--I in the $SU(2)$ case. For the gauge group under consideration,
given an element of the spectrum $\Ab$---i.e., a continuous homomorphism
from $\HAbar$ to the $\star$ algebra of complex numbers---Giles' theorem [10]
provides us with a homomorphism $\hat{H}_{\Ab}$ from the hoop group $\HG$ to
$GL(N)$.

What we need to show is that $H_{\Ab}$ can be so chosen that it takes
values in the given gauge group $G$. We can establish this by
considering the eigen values $\lambda_1,...,\lambda_N$ of
$H_{\Ab}(\at)$. (After all, the characteristic polynomial of the
matrix $H_{\Ab}(\at)$ is expressible directly by the values of $\Ab$
taken on the hoops ${\at},{\at}^2,...,{\at}^n$.)  First, we note from
Eq.\ $(B.1a)$ that in particular $\lambda_i\not=0$, for every $i$.
Thus far, we have used only the algebraic properties of $\Ab$. From
continuity it follows that for every hoop $\at$, $\Tr H_{\Ab}(\at)
\leq N$. Substituting for $\at$ its powers we conclude that
  $$ \| \lambda_1^k\ +\ ...\ \lambda_N^k\| \ \ \leq \ \ N$$
for every integer $k$. This suffices to conclude that
  $$\|\lambda_i\| \ \ =\ \ 1.$$
With this result in hand, we can now use Giles' analysis to conclude that
given $\Ab$ we can find $H_{\Ab}$ which takes values in $U(N)$. If $G$ is
$SU(N)$ then, by the identity $(B.1b)$, we have
  $${\rm det}H_{\Ab}\ \ =\ \ 1,$$
so that the analog of Lemma 3.2 holds.

Combining these results, we have the spectrum theorem: Every element
$\Ab$ of the spectrum $\AGbar$ defines a homomorphism $\hat{H}$ from
the hoop group $\HG$ to the gauge group $G$, and, conversely, every
homomorphism $\hat{H}$ from the hoop group $\HG$ to $G$ defines an
element $\Ab$ of the spectrum $\AGbar$ such that $\Ab (\at) =
\textstyle{1\over N} \Tr \hat{H}(\at)$ for all $\at \in \HG$. Two
homomorphisms $\hat{H}_1$ and $\hat{H}_2$ define the same element of the
spectrum if and only if $\Tr \hat{H}_2= \Tr\hat{H}_1$.

%Some work is required to show that $\hat {H}_1$ and $\hat {H}_2$ are
%conjugate if the above equality holds. This is known to be true
%if for example $\hat {H}_1$ regarded as a representation is not reducible
%[15].

\subsection*{B.3\ \ \ \ \  Cylindrical functions and the induced Haar
measure}
Using the spectrum theorem, we can again associate with any subgroup
$\Sstar\equiv\HG(\bt_1,...\bt_n)$ generated by $n$ independent hoops
$\bt_1, ...\bt_n$, an equivalence relation $\sim$ on $\AGbar$: $\Ab_1
\sim \Ab_2$ iff $\Ab_1(\gt) = \Ab_2(\gt)$ for all ${\gt} \in \Sstar$.
This relation provides us with a family of projections from $\AGbar$
onto the compact manifolds $G^n/{\rm Ad}$ (since, for groups under
consideration, the traces of all elements of a finitely
generated sub-group in the fundamental representation suffice to
characterize the sub-group modulo an overall adjoint map.)
Therefore, as before, we can define cylindrical functions on $\AGbar$
as the pull--backs to $\AGbar$ of continuous functions on $G^n/{\rm
Ad}$. They again form a \C*. Using the fact that traces of products of
group elements suffice to separate points of $G^n/{\rm Ad}$ when $G=
U(N)$ or $G=SU(N)$ [16], it again follows that the \C* of cylindrical
functions is isomorphic with $\HAbar$. Finally, the construction of
Section 4.2 which led us to the definition of the induced Haar measure
goes through step by step. The resulting representation of the \C*
$\HAbar$ is again faithful and diffeomorphism invariant.

\subsection*{B.4\ \ \ \ \  Bundle dependence} In all the constructions
above, we fixed a principal bundle $P(\S, G)$. To conclude, we will
show that the \C* $\HAbar$ and hence its spectrum are {\it
independent} of this choice. First, as noted at the end of Section 3.1,
using the hoop decomposition one
can show that two loops in $\L$ define the same hoop if and only if they
differ by a combination of reparametrization and (an immediate)
retracing of segments for gauge groups under consideration.
Therefore, the hoop groups obtained from any two bundles are the same.

Fix a gauge group $G$ from $\{U(N), SU(N)\}$ and let $P_1(\S, G)$ and
$P_2(\S, G)$ be two principal bundles. Let the corresponding holonomy
$C^\star$--algebras be ${\HAbar}^{(1)}$ and ${\HAbar}^{(2)}$. Using
the spectrum theorem, we know that their spectra are naturally
isomorphic: ${\AGbar}^{(1)} = {\rm Hom}(\HG, G) = {\AGbar}^{(2)}$. We
wish to show that the algebras are themselves naturally isomorphic,
i.e.\ that the map ${\cal I}$ defined by

$${\cal I}\circ(\sum_{i=1}^n a_i T_{\at_i}^{(1)}) = \sum_{i=1}^n
a_i T_{\at_i}^{(2)}\eqno(B.2)$$

is an isomorphism from $\HAbar^{(1)}$ to $\HAbar^{(2)}$. Let us
decompose the given hoops, $\at_1, ...\at_k$, into (strongly)
independent hoops $\bt_1, ...\bt_n$. Let $\Sstar$ be the subgroup of
the hoop group they generate and, as in Section 4.1, let
$\pi^{(i)}(\Sstar)$, $i=1,2$, be the corresponding projection maps
from $\AGbar$ to $G^n/{\rm Ad}$.  From the definition of the maps it
follows that the projections of the two functions in $(B.2)$ to
$G^N/{\rm Ad}$ are in fact equal. Hence, we have:

$$ ||\sum_{i=1}^n a_i T_{\at_i}^{(1)}||_1 = ||\sum_{i=1}^n
a_i T_{\at_i}^{(2)}||_2 , $$

whence it follows that ${\cal I}$ is an isomorphism of $C^\star$
algebras.  Thus, it does not matter which bundle we begin with; we
obtain the same holonomy \C* $\HAbar$. Now, fix a bundle, $P_1$ say,
and consider a connection $A_2$ defined on $P_2$. Note, that the
holonomy map of $A_2$ defines a homomorphism from the
(bundle--independent) hoop group to $G$. Hence, $A_2$ defines also a
point in the spectrum of $\HAbar^{(1)}$.  Therefore, {\it given a
manifold $M$ and a gauge--group $G$ the spectrum $\AGbar$
automatically contains all the connections on all the principal
$G$--bundles over $M$}.

\end{document}